\documentclass[twocolumn,prb]{revtex4}
\usepackage{graphicx}
\usepackage{rotating}
\usepackage{amsmath}
\usepackage{amsfonts}
\usepackage{amssymb}
\usepackage{enumerate}
\usepackage{longtable}
\usepackage{dcolumn}
\usepackage{bm}
\usepackage{latexsym}
\usepackage{epsfig}
\usepackage{subfigure}
\newcommand{\be}{\begin{equation}}
\newcommand{\ee}{\end{equation}}
\newcommand{\bea}{\begin{eqnarray}}
\newcommand{\eea}{\end{eqnarray}}

\begin{document}

\title{Origin of superconductivity in Ni-doped SrFe$_2$As$_2$, an Insight from DMFT}
\author{S. Koley}
\affiliation{Department of Physics, North Eastern Hill University, Shillong, 
Meghalaya, 793022, India}
\begin{abstract}
I describe the issues of the competing orders in normal state of a 
parent and Ni-doped iron pnictide superconductor, $SrFe_2As_2$, using 
LMTO band structure calculation plus multi orbital dynamical mean-field theory.
 Strong, electronic correlations along with minimal antiferromagnetic 
instability create a partially gapped Fermi surface, in qualitative agreement 
with earlier experiments. 
A good semiquantitative similarity in both normal and ordered state with the 
earlier experimental transport data is observed along with suppression 
of magnetic ordering and explained within a correlated, multiorbital viewpoint.
 These results suggest that soft electronic fluctuation mediate 
superconductivity in broad class of parent and underdoped 122 pnictides 
with suppression of magnetic ordering. 
\end{abstract}
\maketitle

\section{Introduction}

\noindent After two decades of the discovery of cuprate high temperature 
superconductivity (SC)\cite{cuprate} a new type of 
high-temperature superconductors has been launched in 
layered iron compounds in 
early 2008\cite{kamijacs}. Addition of these iron compounds 
to the list of ill-understood strongly correlated electronic system (SCES) 
provided a 
new direction of searching high-T$_c$ superconductor other than cuprate based 
systems. Interplay of various interesting effects are almost common in SCES.
Magnetism, superconductivity, and interplay of multiorbital interaction and its 
structure is a novel topic of research in different SCES till date. 
Despite intense efforts a complete understanding of different properties of SCES
is still debatable. 

Discovery of superconductivity at 26 K in fluorine-doped LaFeAsO \cite{kamijacs} has 
triggured a global research to explain the origin of SC in this novel compounds.
 Iron based superconductor has always been surprising because of  
strong local magnetic moment of iron though iron itself is a superconductor 
under high pressure (20 Gpa, 1.8 K) \cite{shimizu}. 
Currently, there is an increased interest in 122 pnictides i.e. AFe$_2$As$_2$ 
(A=Ba, Ca, Sr, Eu) due to pressure induced SC in this compounds \cite{canfield,lonzarich,kotegawa,igawa}. Among them particularly 
SrFe$_2$As$_2$ shows external pressure induced SC at relatively high temperature 
with T$_c$ = 21 K \cite{takahashi}. 
In pressure temperature phase diagram of SrFe$_2$As$_2$ the antiferromagnetic 
phase transition is found at about 198 K and alongwith applied pressure SC 
appears suppressing magnetic ordering\cite{wu}.  
How much strong is the Coulomb interaction in iron pnictides is another 
important query in the field \cite{mazin} with two distinct debated studies: 
the first proposes that the iron pnictides are weakly correlated 
i.e. Coulomb interaction is small compared to the bandwidth while the 
inelastic neutron scattering study posits strong correlation view with the 
observation of drastic reduction of sublattice magnetization\cite{Lynn}.
If it is weakly correlated a standard Fermi 
liquid response in the normal (metallic) state is expected where the spin 
density wave (SDW) is destroyed and dc resistivity should show a T$^2$ 
dependence with a very low residual resitivity at low temperature but 
experimental data exhibits different $T$ dependence\cite{kotegawa}. While weak 
correlation should result in a clean Drude peak with Fermi liquid behavior in 
optical conductivity but perusal of experiments 
show a pseudogap in optical conductivity\cite{okamura} with quasilinear 
temperature 
dependence of resistivity concluding presence of strong correlation in 
SrFe$_2$As$_2$.  
All these experimental data point out that SC arises in this iron pnictide as 
an instability of non-Fermi liquid metallic state.
Earlier theoretical studies also predict incoherent metallic behavior close to a
Mott inslator in other pnictides (LaFeAsO, LaFePO etc.)\cite{boris}.
Such type of theoretical study is not available in Ni doped SrFe$_2$As$_2$.
Though many rich theories have been discovered, the 
mechanism of superconducting transition is remaining as a controversial subject.
How from an incoherent metal the SC can arise in parent material with 
isoelectronic substituion will be aim of this article.

Theoretically explaining mechanism of 
high T$_c$ superconductivity lies on identifying features that are directly 
tied to high T$_c$ among complexity of different physical phenomena. 
So in this paper it is shown that the 122 iron pnictides in doped or 
parent phase can be a new class of materials to test the superconductivity 
. I have compared the electronic properties of Ni doped and parent SrFe$_2$As$_2$ obtained from 
DMFT with earlier experiment and established the idea of electronic correlation behind different 
transition in its doping temperature phase diagram. Since parent 122 pnictide 
is doped with Ni, the magnetic interactions become weak. Electronic structure 
near Fermi level is mainly from t$_{2g}$ orbitals, which form electron pockets 
in Fermi surface and superconductivity emerges as a consequence of weak 
of magnetic order. 

In following sections, I calculated the DMFT spectral DOS and optical 
conductivity of parent and Ni doped SrFe$_2$As$_2$. I will show how orbital 
selective Mott transtition underpins the properties of this 122 pnictide. In 
particular I will present specific non-FL feature behind phase transition of 
the system. 

\section{Method}
\begin{figure}
\centering
\includegraphics[angle=270,width=\columnwidth]{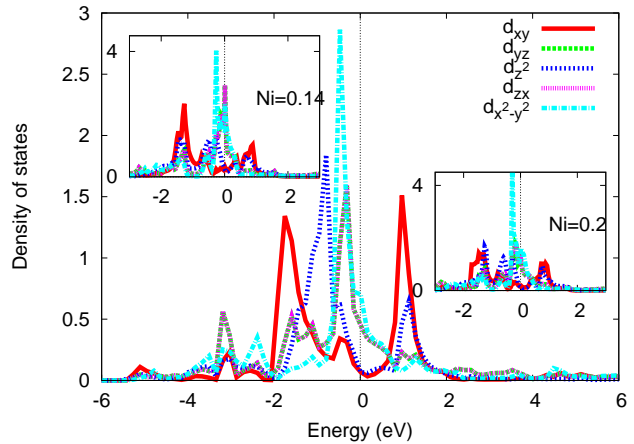}
\caption{(Color Online)LMTO density of states of parent $SrFe_2As_2$ in the 
orthorhombic structure. Left and right inset is for doped $SrFe_2As_2$ with 
Ni=0.14 and Ni=0.2.)}
\label{fig1}
\end{figure}

The one electron band structure of SrFe$_2$As$_2$ was calculated using the 
linear muffin-tin orbital (LMTO) scheme \cite{andersen} in the atomic sphere 
approximation (ASA) as a function of doping concentration in 
the orthorhombic as well as tetragonal phases. Lattice parameter 
inputs of LMTO are taken from earlier experimentally determined a, b, c, and 
$z_{As}$ as a function of doping and temperature\cite{saha}. In SrFe$_2$As$_2$ 
the effective electronic states include the carriers in the FeAs 
layers. In Fig.\ref{fig1} I show the orbital dependent DOS of five Fe-d bands 
which have significant weight at Fermi energy (E$_F$) in both parent and 
doped Sr122
system. The density of states (DOS) shows that among the five d-orbitals $xy$ 
and $z^2$ band is almost gapped at the Fermi energy and other bands have large 
weight at the $E_F$. This noninteracting DOS in doped $SrFe_2As_2$ changes 
at the Fermi energy due to redistribution of electrons with doping. 
Theoretically the system is doped using supercell approach  
and paramagnetic calculations are carried out with I4/mmm 
space group symmetry (No. 139) in tetragonal phase and Fmmm space group 
(No. 69) in antiferromagnetic orthorhombic phase respectively. 
The noninteracting part of the Hamiltonian is given by
\begin{equation}
H_0=\sum_{k,a,\sigma}\epsilon_{k,a}c^{\dagger}_{k,a,\sigma}c_{k,a,\sigma}
\end{equation}
where $\epsilon_{k,a}$ is band energy for the five Fe-d bands. The crystal 
field with 
$S_4$ symmetry creates inter-orbital splitting in iron-pnictides (Fe-pn).
As discussed earlier the strong coupling view agrees well with quasi-local spin
fluctuations associated with Mott physics and subsequent ARPES \cite{Yi} and 
STM\cite{arham} studies at normal state.
Theoretically it is well established that in a strongly correlated system LDA 
(local density approximation) is unable to describe excited states while 
LDA+DMFT(Dynamical mean field theory) has been successful in describing physical
 properties of various correlated electronic structures.
The multi orbital iterated perturbation theory (MO-IPT) is used as an impurity 
solver in DMFT: though not exact, it is a computationally fast and effective 
solver, and has been proven to work very well in real multi-band systems 
throughout all temperature range\cite{at1,at2,at3}. I choose U= 0.5-0.7 eV and 
U$^\prime$= 0.2 eV as 
intra and inter-orbital Coulomb interaction as appropriate (for good description of earlier experiments) parameters and J$_H$ is determined from the relation 
U$^\prime$=U-2J$_H$. 
The interaction part of the Hamiltonian is given by, 

\begin{equation}
H_{int}=U\sum_{i,a}n_{ia\uparrow}n_{ia\downarrow} + U'\sum_{i,a,b,\sigma,\sigma'}n_{ia\sigma}n_{ib\sigma'}-J_H\sum_{i,a,b}S_{ia}.S_{ib}
\end{equation}

Further in pnictides the relevant phonon mode which couples electrons is given 
by $H_{el-l}=g\sum_i(A_i+A_i^\dagger)(c_{ia}^\dagger c_{ib}+ h.c.)$ (g is taken 0.01 after 
checking all possible realistic values to get a good description of 
experiments).
Given these I have followed earlier procedure \cite{at1,at2,at3} to incorporate 
interaction effects in the five-band model
above within the DMFT.

\section{DMFT Results}
In Fig.2a I show DMFT local density of states for different temperature. 
A significant change is noticed in the sharpness of the DMFT spectral function 
for the bands which are very close to the Fermi energy as the temperature cross 200K. 
The dynamical correlations lead to change in spectral weight over an 
higher energy scale. At 200 K spectral DOS reveals 
finite energy gap as shown in Fig.2a. Thus these result strongly suggest that 
AFM correlation sets in orbital selective Mott transition (OSMT) in 122 
pnictide (See supplementary information for smaller range near Fermi energy). The DMFT 
studies also suggest that the selective gaps will also appear in band structure
at strong limit of the AFM correlations, while a Mott transition is caused by 
strong Coulomb interaction notwithstanding the magnitude of magnetic 
correlations. In parent $SrFe_2As_2$ the OSMT coincides with the AFM transition 
which thus establishes strong magnetic order in the system.  
Further there is no sign of FL quasiparticle signature in the low energy 
spectra, which drive orbital selective Mott transition in the d-band.
Fig.2b presents spectral functions of $SrFe_{1.86}Ni_{0.14}As_2$ as a funtion of 
temperature. With increasing x, a and c-axis lattice parameters are changed.
Increasing Ni concentration 
change the spectral DOS in a significant way. The DMFT spectral function becomes
 more coherent which indicates the reduced correlation. Finite spectral weight 
at the Fermi energy represents the metallic nature of the doped 122 
pnictide, as the weight decreases with increasing $T$ the system evolves to 
incoherent metal, the self energy also shows the deviation from $\omega ^2$ 
behaviour. As the doping is increased from x=0.14 to x=0.2 a 
small pseudogap is formed in one of the d-band (Fig.2c) which also predicts 
to set an ordering 
at low temperature where the electronic correlation is suppressed and 
incoherent fluctuations are supposed to be condensed due to lowering in 
temperature resulting in the development of ordered phase with 
increasing doping.

\begin{figure}
\centering
(a)
\includegraphics[angle=270,width=0.8\columnwidth]{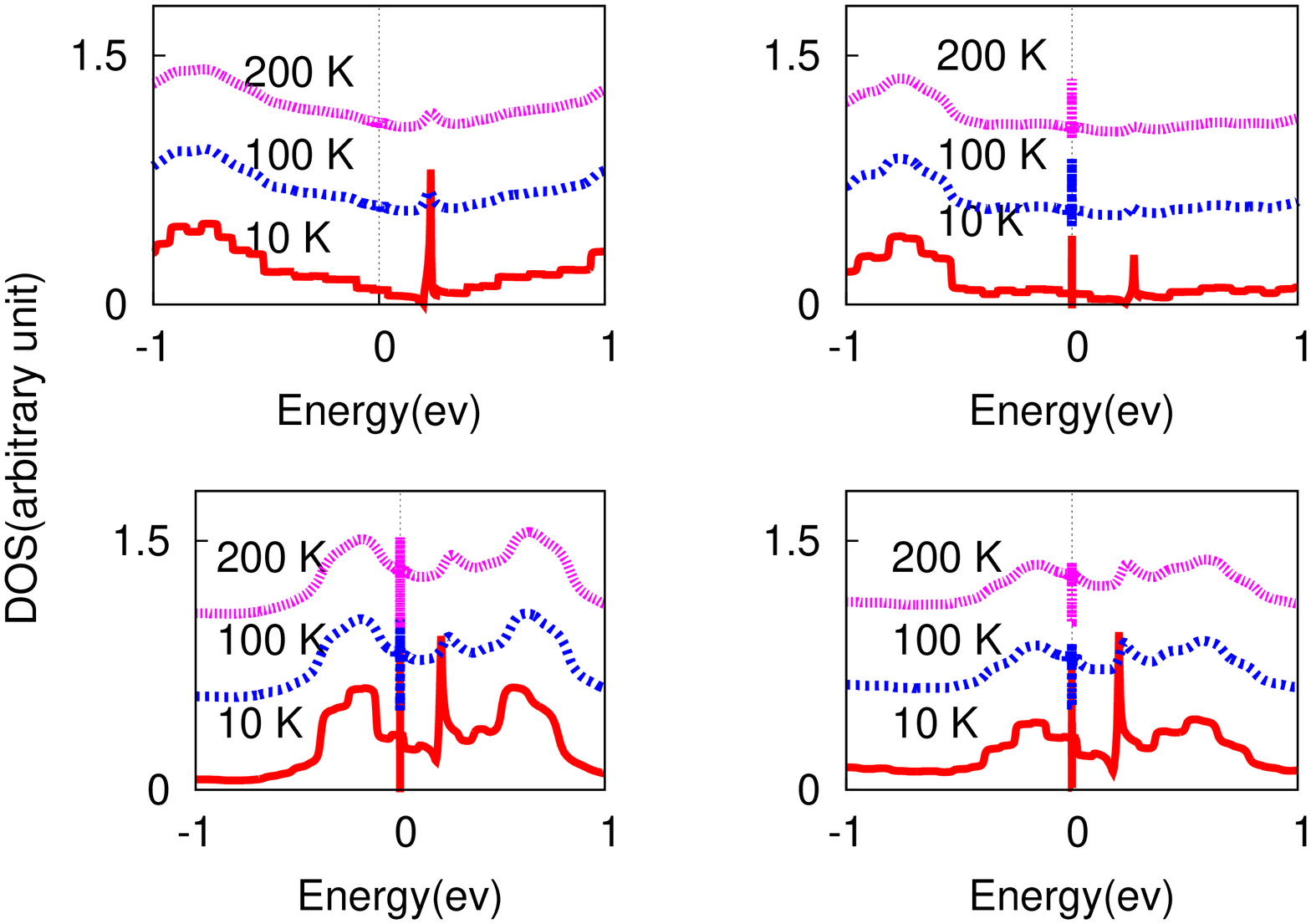}

(b)
\includegraphics[angle=270,width=0.8\columnwidth]{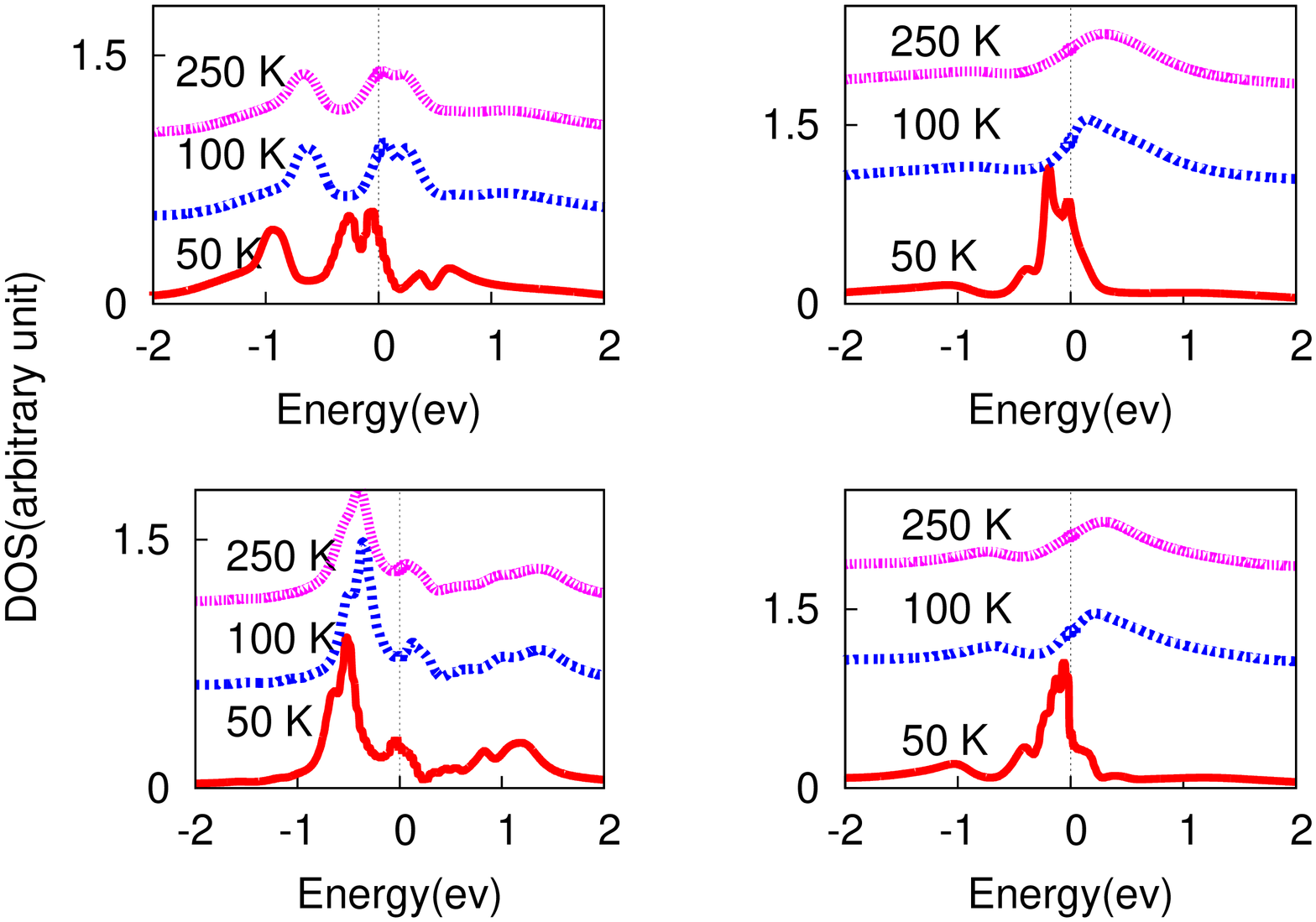}

(c)
\includegraphics[angle=270,width=0.8\columnwidth]{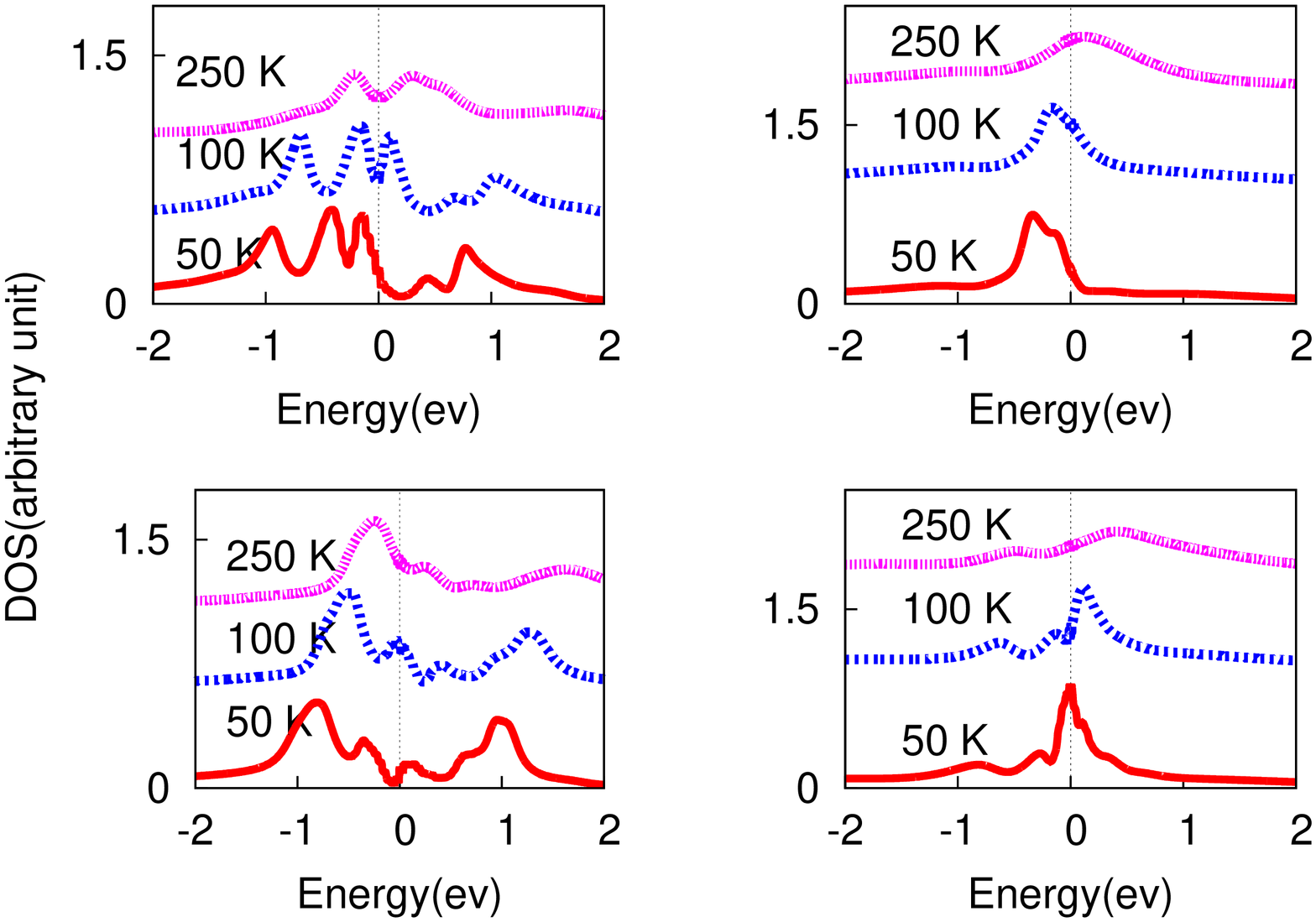}
\caption{(Color Online)(a) DMFT density of states of parent $SrFe_2As_2$ at 
different temperature. In upper pannel $yz$, $x^2-y^2$ and 
in lower panel $xy$ and $z^2$ DOS is shown. Sharp pole at $E_F$ manifests 
orbital selective Mott localization. DMFT density of states of 
doped $SrFe_2As_2$ with (b)Ni=0.14 and (c)Ni=0.2. In upper pannel $yz$, 
$x^2-y^2$ and in lower panel $xy$ and $z^2$ DOS is shown.}
\label{fig2}
\end{figure}

Normal state transport also can be explained from this theory. 
I next compute the optical conductivity using DMFT which is facilitated by the 
finding \cite{tomczak}that irreducible vertex corrections in the Bethe-Salpeter 
equations for conductivities are negligible and can be ignored to a very good 
approximation. Fig.3 shows the optical conductivity ($\sigma(\omega)$) spectra 
obtained from DMFT. In the parent Sr-122 system $\sigma(\omega)$ rises rapidly 
with decreasing frequency and almost featureless above 0.5 eV. Since I have 
taken bands closest to Fermi energy, higher energy resemblance with earlier 
experiments is not expected. The Drude response of free carriers is discernible
 in low energy optical conductivity up to lowest temperature. 
With cooling below 200 K  
optical conductivity at lower energy is depleted and with further cooling it 
shows a gap-like structure. Further a shoulder appears near 0.5 eV and 
the Drude response shrinks with decreasing $T$. The formation of gap in 
optical conductivity has been observed earlier also \cite{hu,Wu,okamura}.  
These results conclude that Fermi surface is gapped at few areas 
of the Brillouin zone due to the AFM ordering. The DMFT optical 
conductivity with two different Ni doping level is presented in Fig.3b and 
Fig.3c. 
The spectral weight transfer with cooling is still discernible in 
$\sigma(\omega)$ but a large reduction is 
observed at 100 K, rather than below 200 K as in parent Sr122 which may 
be related to magnetic ordering transition. This result corresponds the decrease
 in ordering temperature due to Ni doping. However with doping the shoulder like 
feature is found below 100 K in the energy range 0.1 eV which is 
smaller than parent Sr122. Increase in the Ni doping $\sigma(\omega)$ follows 
the same trend overall. The spectral weight transfer becomes smaller with 
narrower Drude response and the shoulder like feature also become insignificant 
throughout the temperature range except 10 K. $\sigma(\omega)$ is identified 
almost flat 
in every temperature. Based on these considerations and the optical conductivity
 results it can be inferred that magnetic ordering transition is suppressed due 
to Ni doping and the electron lattice interaction plays a significant role 
in suppression of magnetic ordering.

\begin{figure}
\centering
(a)
\includegraphics[angle=270,width=0.8\columnwidth]{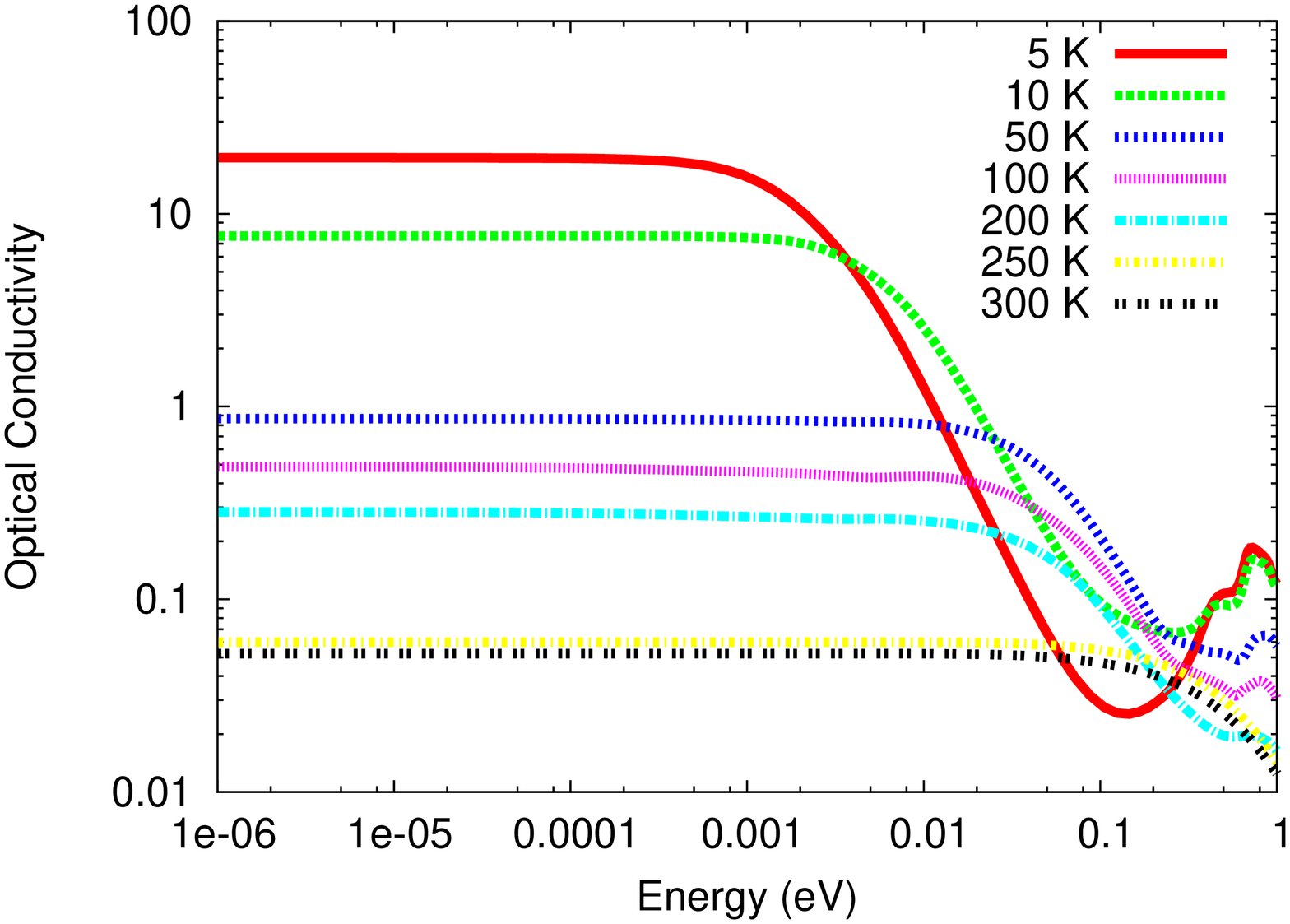}

(b)
\includegraphics[angle=270,width=0.8\columnwidth]{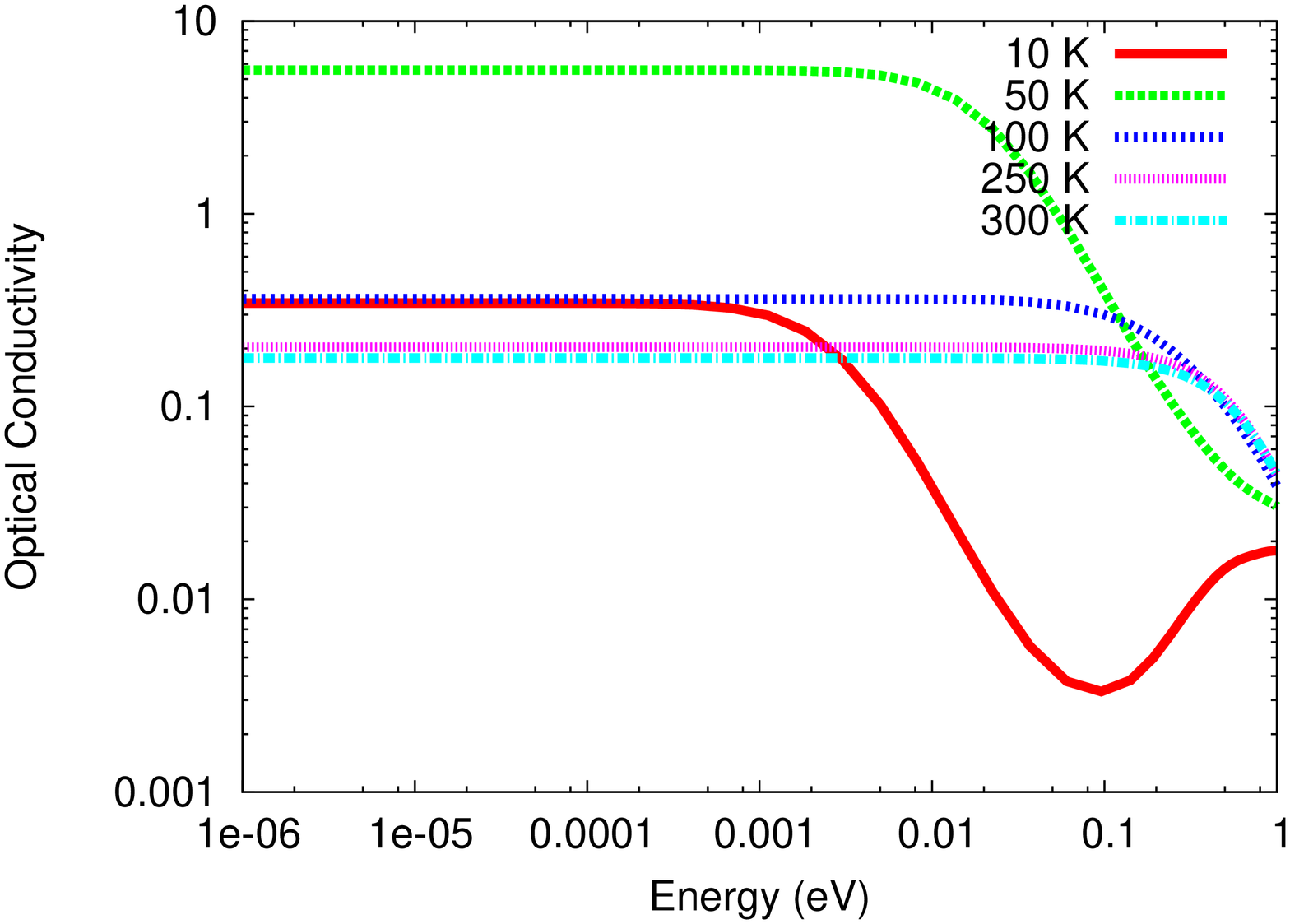}

(c)
\includegraphics[angle=270,width=0.8\columnwidth]{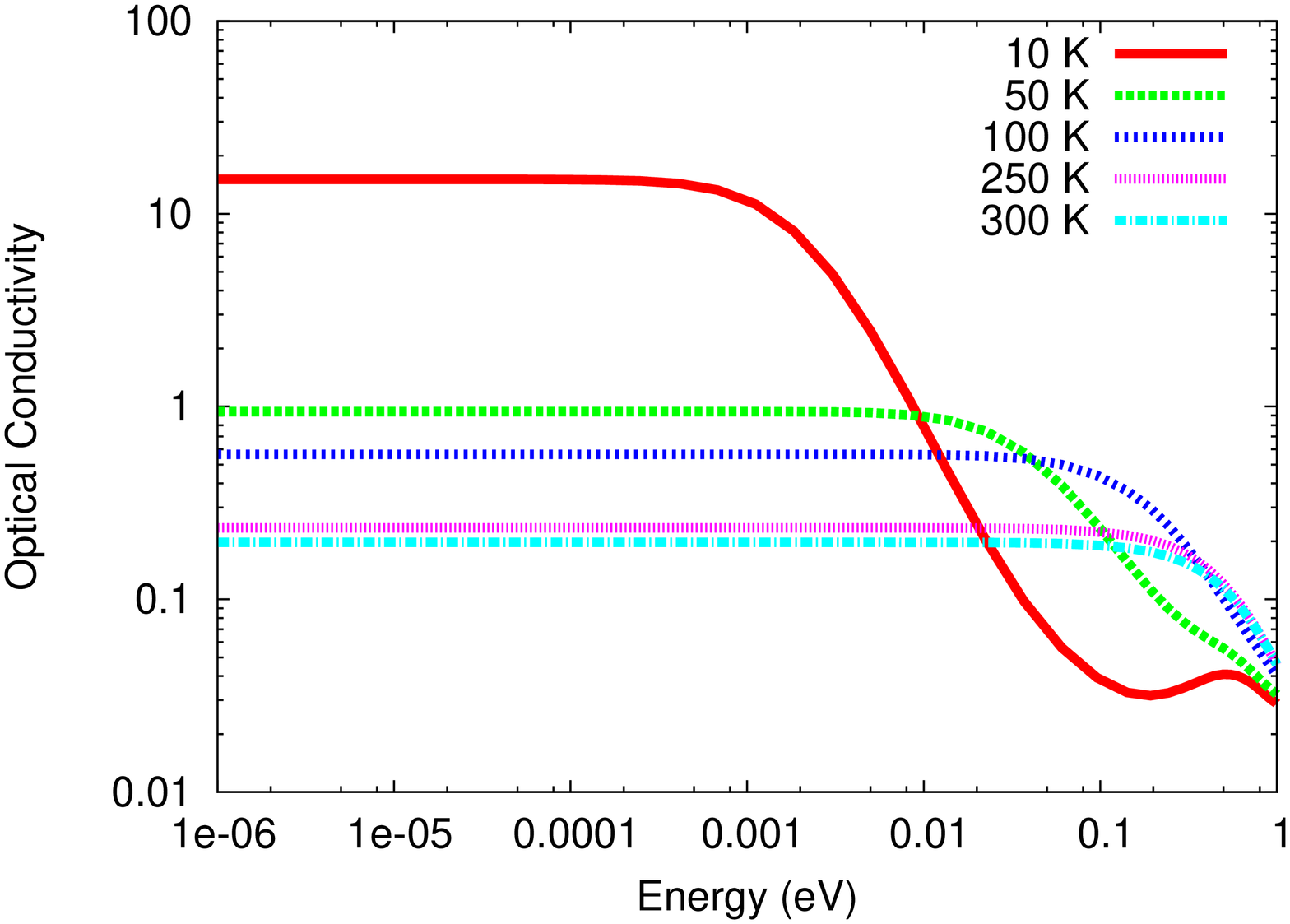}
\caption{(Color Online)(a) DMFT optical conductivity (plotted in logscale) of 
parent $SrFe_2As_2$ at different temperature. DMFT optical conductivity of 
doped $SrFe_2As_2$ with (b)Ni=0.14 and (c)Ni=0.2.}
\label{fig3}
\end{figure}

Fig.4a presents the electrical resistivity $\rho(T)$ of parent and doped 
$SrFe_2As_2$. The dc resistivity is computed here using the Kubo formalism 
in DMFT.
 Interestingly $\rho(T)$ for Sr122 manifests metallic behavior, 
decreasing with T 
from 300 K and linear in $T$ till around 200 K. At around 200 K, $\rho(T)$ 
manifests a sharp kink, which is exactly the temperature where AFM ordering sets 
in. With Ni doping the anomaly in resistivity becomes less distinct and is 
defined by change in the slope of resistivity. Finally in x=0.2 no such anomaly 
is found. The sharp reduction of $\rho(T)$ in parent material is changing 
behavior with increased Ni concentration as it is shifted to lower 
temperature 
in full accord with experimental data \cite{saha}. This type of behavior in 
resistivity has already been observed in other doped 122 pnictides \cite{rosner, 
canfield}. Since this anomaly is associated with a magneto structural transition
 so it can be due to reduced incoherent scattering at the onset of magnetic 
ordering and structural transition associated changes in carrier concentration. 
Also this concludes dominant role of electronic scattering in resistivity. 
To find out actual $T$ dependence of resistivity if $\rho(T)$ is fitted with 
the expression $\rho(T) = \rho_0+AT^n$ for the parent Sr122. In the 
AFM ordered region it is found that the value of n remains around 2.7, this fact
 is rather unusual and it represents quantum fluctuation around that region.

\begin{figure}
\centering
(a)
\includegraphics[angle=270,width=0.8\columnwidth]{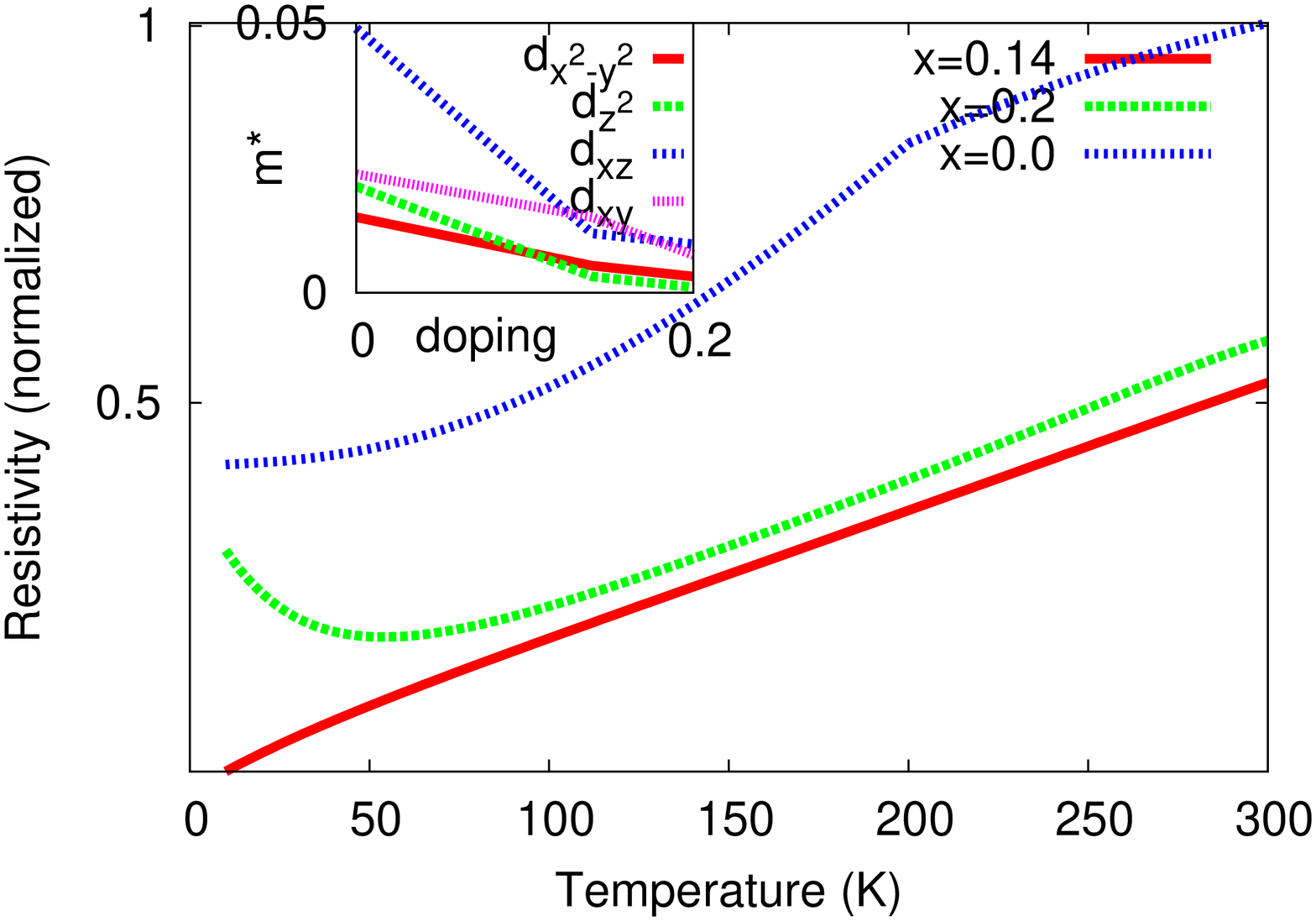}

(b)
\includegraphics[angle=270,width=0.8\columnwidth]{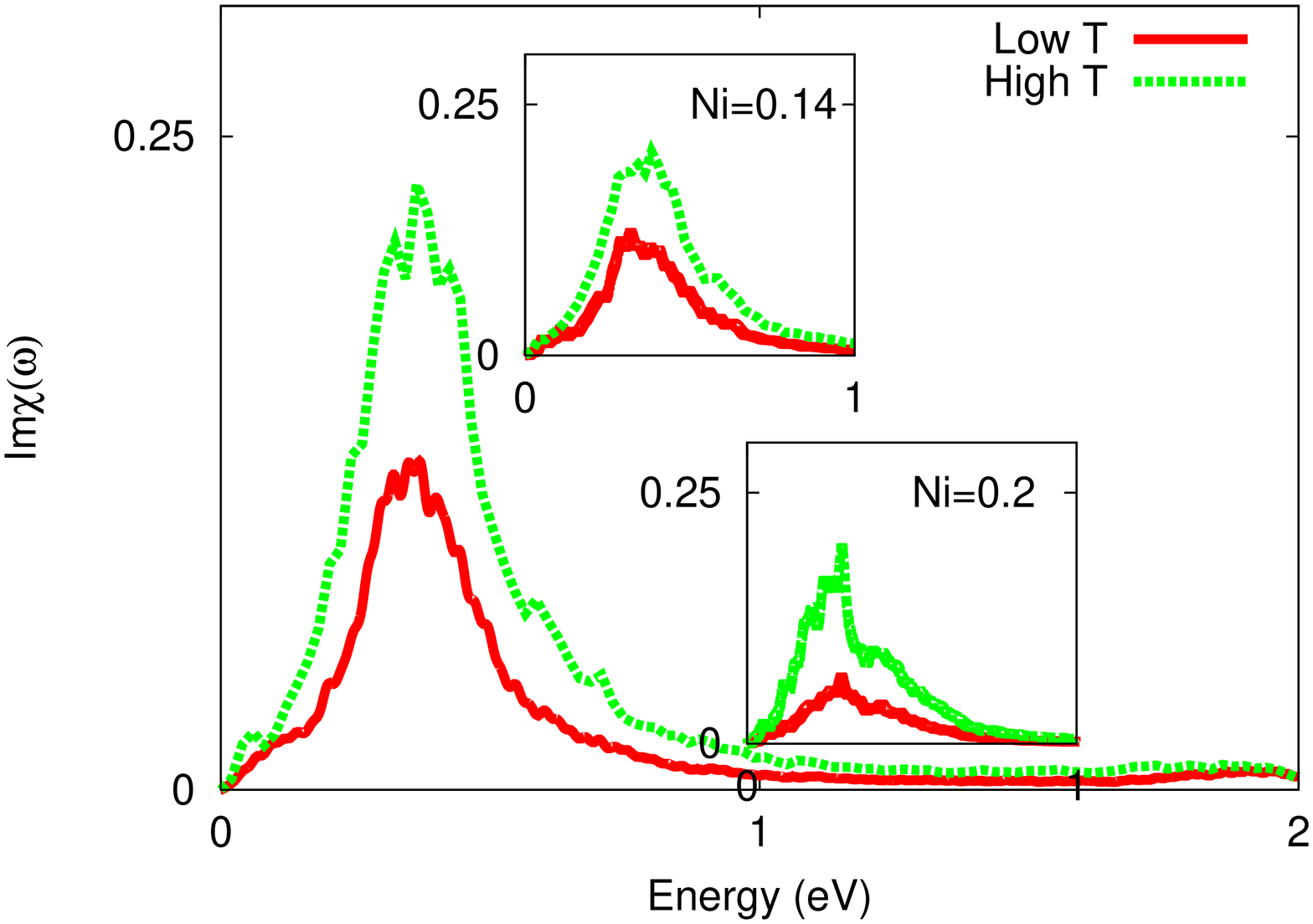}
\caption{(Color Online) (a) DMFT resistivity of parent $SrFe_2As_2$ at
different temperature and doped $SrFe_2As_2$ for Ni=0.14 and Ni=0.2. Inset 
shows change in effective mass for different orbital with doping in the normal 
state. (b) Imaginary part of dynamic susceptibility showing magnetic fluctuation
 in high and low temperature phase. Inset shows the same for two different doping.}
\label{fig4}
\end{figure}

To investigate the effect of doping on the magnetic order and fluctuating moments,
I computed the dynamic magnetic susceptibility $\chi(\omega)$, which 
estimates the temporal and spatial dependence of fluctuating magnetic moments. 
In Fig.4b the $\chi(\omega)$ varying with real frequency is presented for both 
undoped and doped phases. The Matsubara frequency dependent local dynamic 
susceptibility is calculated from DMFT first and then it is used to get $\chi(\omega)$ as a function of real frequency using maximum entropy method. 
Sharp peak is noticed in $\chi(\omega)$ at low energy which dies out with 
increasing energy. The peak here evidents 
large fluctuating moment which is very obvious in the high temperature-phase. 
Whereas a substantial drop in local moment in the low temperature ordered phase 
makes the susceptibility diminishing at high energy. Similarly in
 the doped phase also the peak height shrinks which establishes reduction of 
local moment oscillation in the ordered and doped 122 pnictide.

Now to further investigate about the ordering in the doped phase I calculated 
the degree of correlation. To know about that I will present effective mass
$(m*=1/Z)$, where $1/Z =(1−\delta\Sigma/\delta\omega)_\omega=0$. In a Landau 
Fermi liquid $Z$ is denoted as the quasiparticle weight, for a
non-interacting system $Z$ becomes 1, and for a strongly correlated system it 
is very much smaller than 1 and near about zero for an insulator. 
I have computed effective mass for all the five d orbitals and plotted them with 
change in doping in Fig. 4a inset. A subtantial drop in the $m^*$ for all 
Fe-d orbitals is found while varying doping in Sr122. It is found that 
the $d_{z^2}, d_{x^2-y^2}$ orbitals are less correlated and the 
$t_{2g}$ orbitals ($d_{xz}, d_{yz},$ and $d_{xy}$) are more
correlated in parent Sr122. The effect of doping on effective mass is observed 
to undergo a large change in the $d_{xz}$ orbital. This iron pnictide in undoped
 phase exhibit AFM ordering which is due to strong effective NNN (next nearest neighbour) exchange coupling between two iron atoms. This NNN coupling must 
originate from coupling between iron t$_{2g}$ d-orbitals and the anion 
p-oribitals and superconducting pairing with extended s-wave symmetry is 
expected to stem from this AFM interaction. The electronic structure of t$_2g$ 
orbitals near Fermi level generates the AFM interaction which is expected to 
be the key of phase transition in the 122 pnictide. The change in effective 
mass due to doping in t$_{2g}$ d-orbitals puts another proof of development of 
superconductivity with suppressed magnetic ordering.

Here, I presented a theoretical study of parent and doped $SrFe_2As_2$ to 
investigate the magnetism, superconductivity following suppressed magnetic 
ordering
and the relation of crystal structure and charge doping to the different phases.
 Started with LMTO electronic structure calculations the study is centred on 
doping dependent changes in transport studies and photoemission spectras 
for Sr-122. Since the electronic structure is quite similar to the other 
pnictide families so the physical properties can be predicted and this approach
provides insight into many questions. 
Since accurate description of the interaction of the iron arsenide layer and 
interlayer distance from geometry optimized density functional calculation 
is still debatable I have taken experimentally determined lattice parameters 
as function of temperature and doping. Electronic structure calculation 
essentially represent the signature of structural and magnetic order related 
changes with increased doping or temperature but the presence of correlation 
in the system is treated within DMFT. It is found that the structural 
transition in the 122 compounds is intrinsically connected with 
antiferromagnetic 
transition. I noticed suppression of the magnetic ordering with Ni doping 
which is in excellent accord with earlier experiments. Moreover theoretical 
ARPES studies (See Supplementary Information) suggest FS is gapped only in few direction. Doping dependence of 
the parent Sr122 in this theory follows correct trends as in experiments but a 
quantitative comparision will call for direct treatment of the influence of 
disorder in structure and magnetism. As discussed in the text the magnetism 
and therefore the superconducting order is sensitive to the position of As and 
it crucially depends on the structure and interaction of iron arsenic layer.
Thus very good quantitative agreement with an entire range of spectral and 
transport data for this 122 pnictide and qualitative rationalization of 
structural features from a single theoretical calculation supports the novel 
strong coupling picture.
So all the results points towards a strong coupling superconductivity arising from an incoherent non-Fermi liquid metal.

From these findings one can easily predict about mechanism of superconductivity 
under pressure or doping. Finite pressure or doping increases the band overlap 
and thus leads to redistribution of electrons and holes which is found in the 
noninteracting DOS here also. Redistribution of electronic energies weakens 
magnetic order which can form the critical electronic fluctuationand thus lead 
to an instability to superconductivity with a finite gap in the spectrum. 
However, implication form the incoherent paramagnetic metal in the normal state 
suggests that superconductivity can arise form incoherence via two particle 
instability by solving $H=H_{el}+H_{res}$ in the pair channel like earlier \cite
{at1}, where decoupling the intersite (interband) interaction in a generalised 
HF sense it can be derived $H_{res}^{HF}=
p\sum_{a,b,k,\sigma,\sigma'}(\langle
c_{k,a,\sigma}^{\dagger}c_{-k,b,\sigma'}^{\dagger}\rangle c_{-k,b,\sigma'}c_{k,a,\sigma}+ h.c.)$. Such term arises at second order from one-electron inter-band 
term, when the one-electron spectral function is incoherent and the superconducting order parameter 
can be calculated from $\Delta_{ab}(k) \propto \langle c_{k,a,\sigma}^{\dagger}c_{-k,b,\sigma'}^{\dagger}\rangle$ which yields multiband spin-singlet SC. 
The electron lattice interaction will further increase the interaction.
 Considering the lattice symmetry and including NNN AFM coupling $\Delta_{ab}(k)
$ can be explicitly written as $\Delta_{ab}(k)=\Delta(2cos(\frac{\sqrt 3}{2}k_x)cos(k_y/2)+cosk_y)$. The earlier $s\pm$ proposal for superconductivity was 
derived from weak-coupling instability of an itinerant Fermi liquid. The present
 extended s wave idea is a generalization of the earlier proposal.
A crucially significant outcome of my work is magnetic order induced 
changes coincident with the structural transition and interband proximity effect
giving possibility of c axis nodal structure in 122 pnictides. Moreover NNN 
antiferromagnetic superexchange 
coupling puts constraint on finite and k-dependent $\Delta_{ab}(k)$. 
Since an OSMT is found the corresponding Fermi surface will undergo a 
drastic change. Finally unconventional superconductivity arises as an 
instability of incoherent normal state. In summarizing, this 
LDA+DMFT analysis explains a large range of experimental observations and 
discovers a link between structural and electronic fluctuations. A range of 
unusual responses in Sr122 is clarified from this theory. In the selective Mott 
 transition vanishing of Fermi surface in some portions of Brillouin zone 
corresponds to an divergence of effective mass. Thus, altogether these strongly supports the proposal of unconventional SC phase from an extended s-wave pairing. Similar results can be expected for same parent structure.

\section{ACKNOWLEDGEMENT}
S.K. acknowledges support from DST women scientist grant SR/WOS-A/PM-80/2016(G).

\pagebreak
\section{Supplementary Information}
\subsection{Angle resolved photoemission within DMFT(IPT)}

It is needed to inquire the microscopic origin of this ordering. Below 
$T=200K$ three d-bands are Mott localized due to interaction (but there is 
lack of clean gap due to inequal changes in Fermi surface, which completely 
remove the idea of FS nesting induced order). If this theory is to be credible,
 then other observations also must go hand in hand with experiments. In Fig.s1
and Fig.s2 I 
show DMFT one particle spectral function $A(k,\omega)=-ImG(k,\omega)/\pi$ which 
also reflects renormalized band dispersion $E_{k,a}=\epsilon_{k,a}+Re\Sigma_
{k,a}(\epsilon)$. Though in-detail ARPES study is not available in literature 
for Ni doping but photoemission spectra for parent Sr122 can be compared with 
earlier results. While LDA with static Hartree Fock can show agreement with band
 dispersion, simultaneous explanation of ARPES will test the theory. For a high 
temperature dynamically fluctuating liquid ARPES is presumed to reveal 
broad 
continuum like features without Landau Fermi liquid quasiparticle peaks. 
However the DMFT results here provide a good description of extant ARPES 
dispersion upto high energies. Particularly they unveil the gap features in the 
ordered state. Fig.s2 shows the normal state map of the photoemission intensity 
from $\Gamma$ to $M$ direction. Fig.s1 illustrate the photoemmission 
intensity (PES) energy distribution curves (EDC) across $\Gamma$ to $M$ 
direction in the antiferromagnetic ordered state. At low $T$ ($T$=5 K, Fig.s1a-s1d) two 
bands can be identified to cross Fermi Energy providing two holelike Fermi 
surface portion around $\Gamma$ point of the Brillouin zone, in accord with 
earlier band structre calculations. Additionaly there is a feature at around 
-0.2 eV, which can be identified as the valence band backfolded and hybridize 
with hole-like bands to open energy gaps near Fermi energy. Fig.s1e-Fig.s1h and 
Fig.s2 shows 
the ARPES intensity at two higher temperatures 200 K and 300 K. The EDCs are 
dominated by a strong feature dispersing towards M point. EDCs at 200 K reveals 
that the hole like bands moves towards $E_F$. From the ARPES at 
200 K a new peak can be identified near $E_F$ originating from coupling 
to phonons which disappears with increasing temperature however at high $T$ an 
extra low energy feature is resolved which remains almost dispersionless.  
Fig.s3 and Fig.s4 show ARPES intensity along $\Gamma -M$ direction of Ni doped Sr122. 
The calculated EDCs are very similar in two doping state. However details of 
the electronic structure around $E_F$ for doped and undoped state were 
sensitive to this parameter. At high temperatures the dispersive features of 
the EDCs near E$_F$ is considerably suppressed due to orbital and spin 
fluctuations in the paramagnetic metal phase. As identified from the EDC plot 
at low temperature the spectral weight near Fermi energy is increased 
in the 
ordered state in comparison to paramagnetic phase. This finding infers that 
due to magnetic ordering transition the intensity suppression caused by the orbital
 and spin fluctuations is eliminated.

\begin{figure*}
(a)
\includegraphics[angle=270,width=0.8\columnwidth]{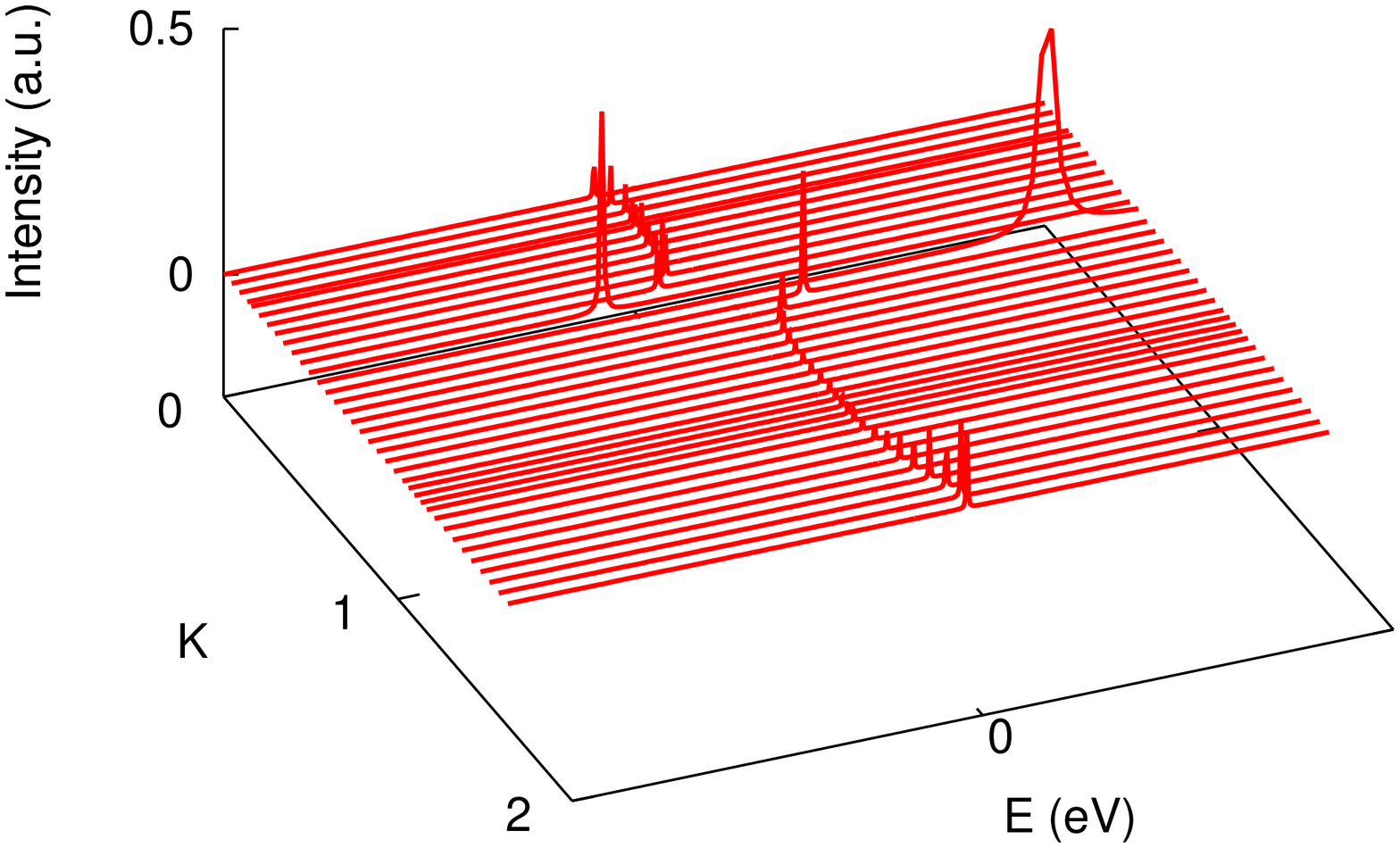}
(b)
\includegraphics[angle=270,width=0.8\columnwidth]{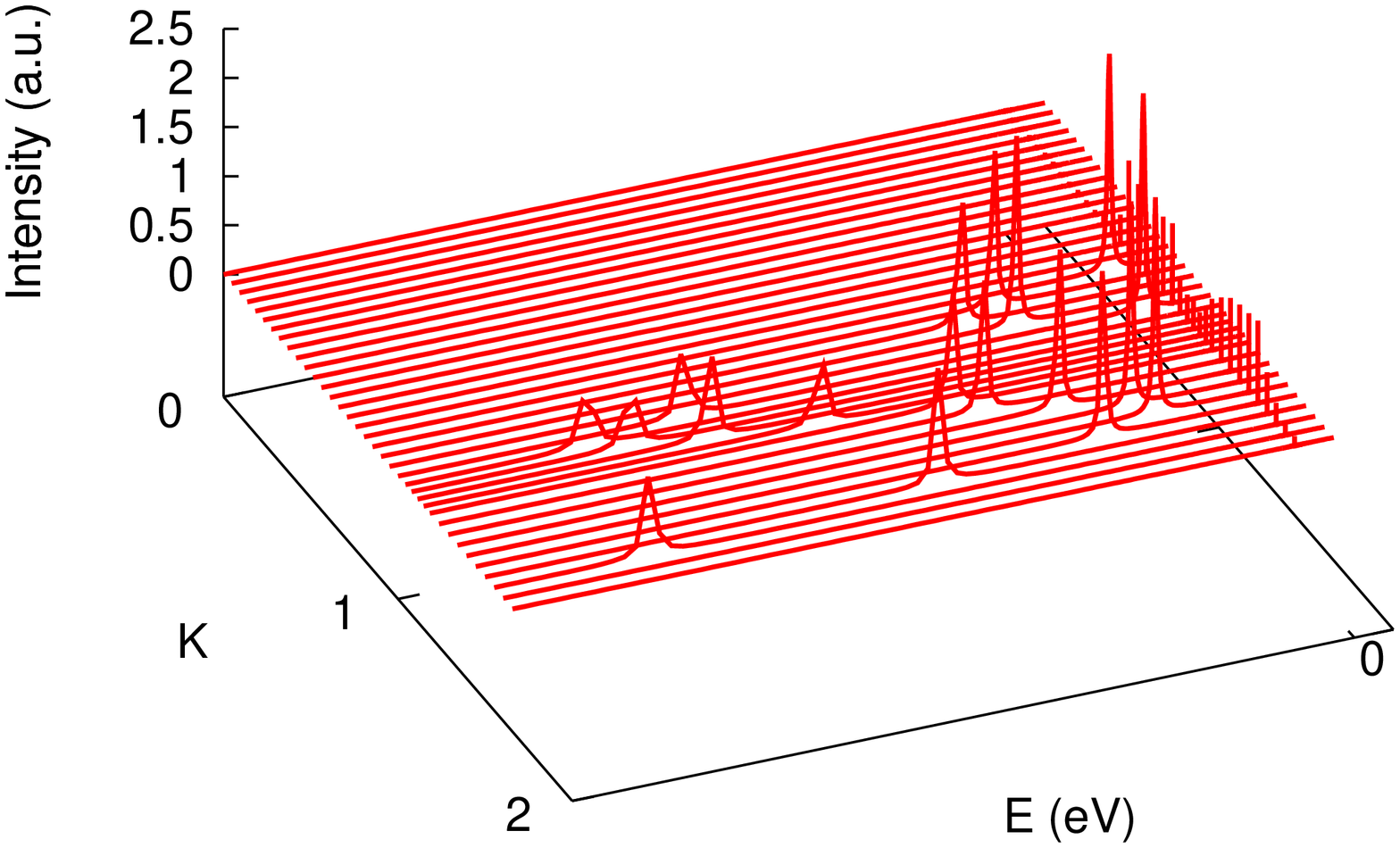}

(c)
\includegraphics[angle=270,width=0.8\columnwidth]{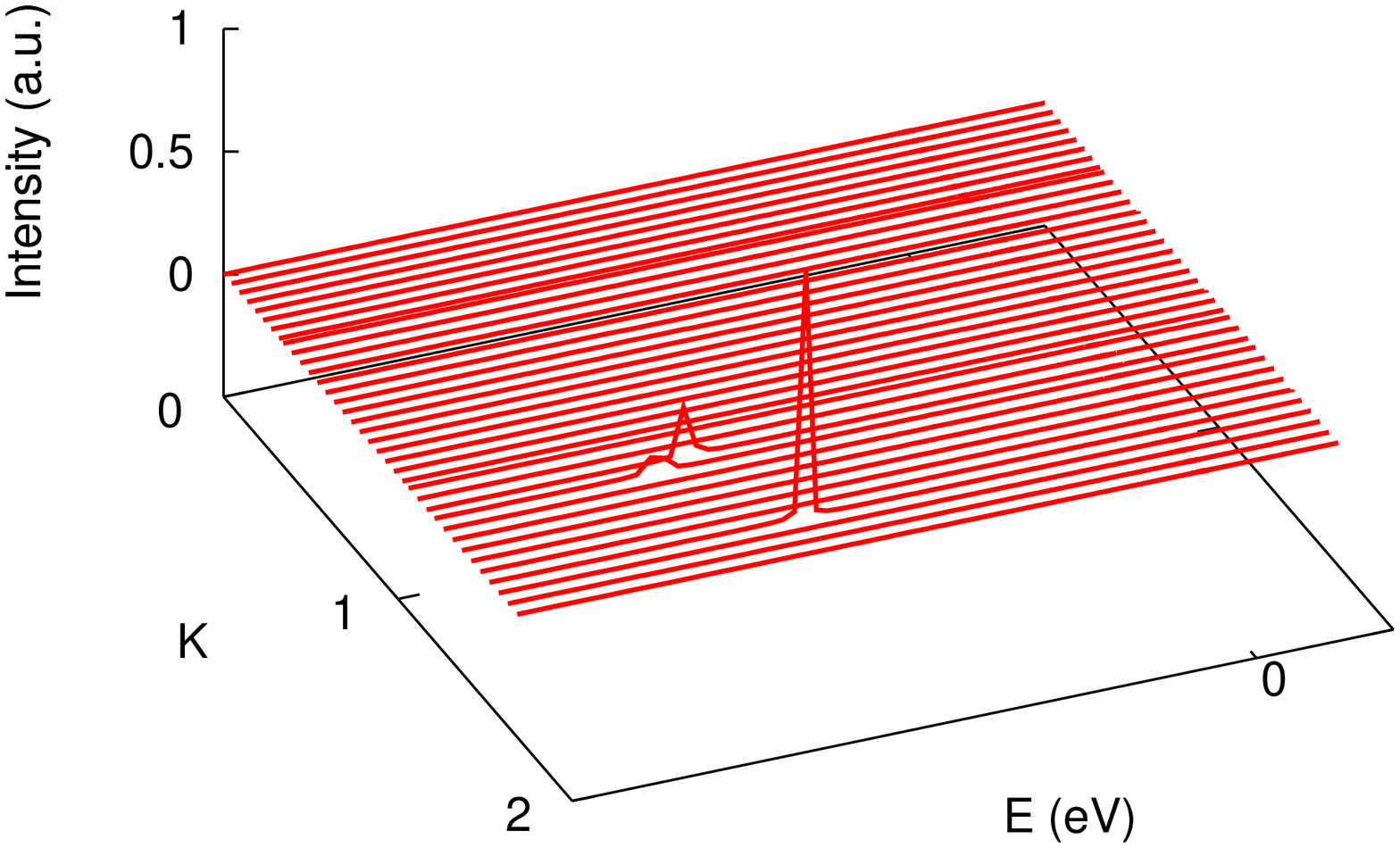}
(d)
\includegraphics[angle=270,width=0.8\columnwidth]{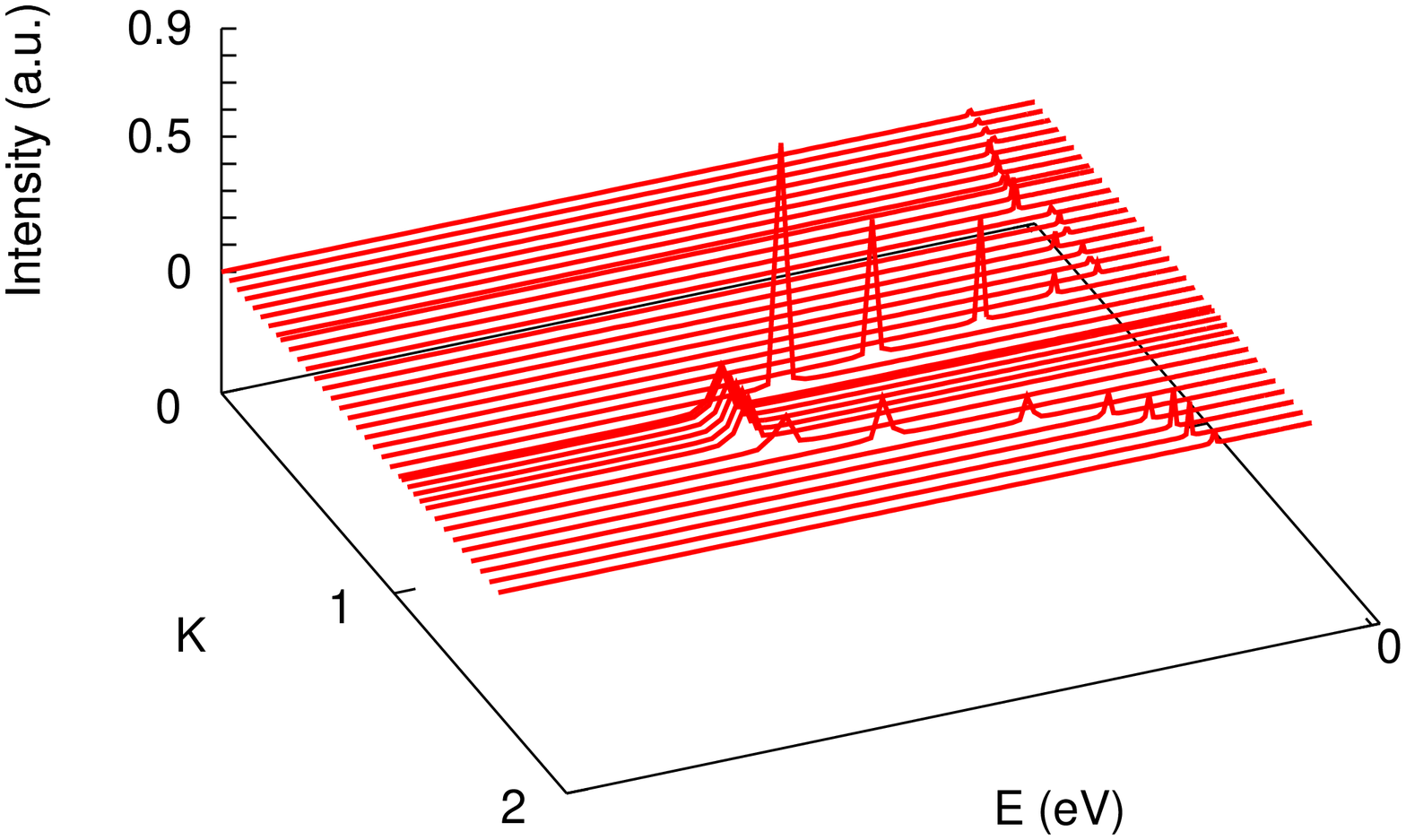}

(e)
\includegraphics[angle=270,width=0.8\columnwidth]{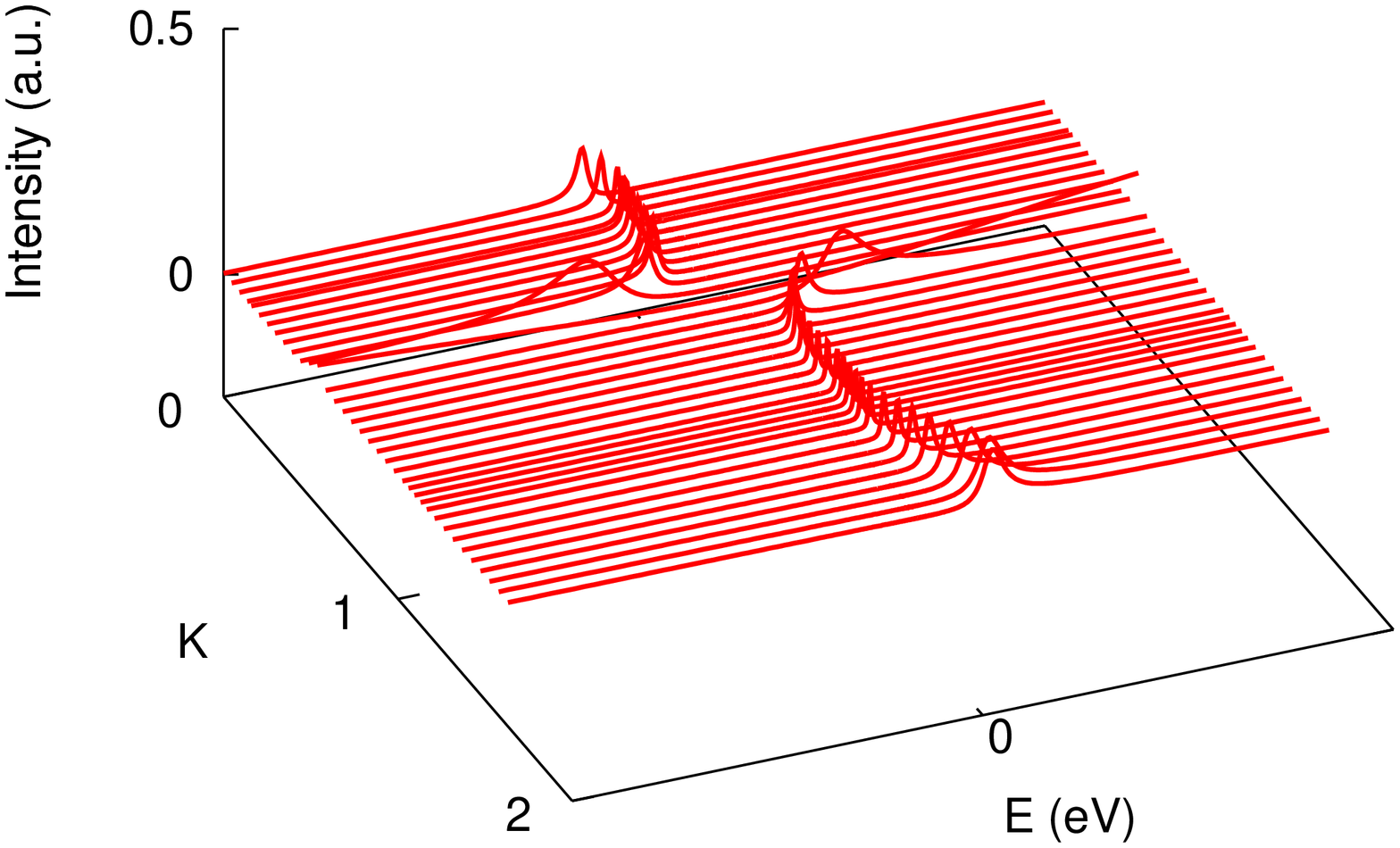}
(f)
\includegraphics[angle=270,width=0.8\columnwidth]{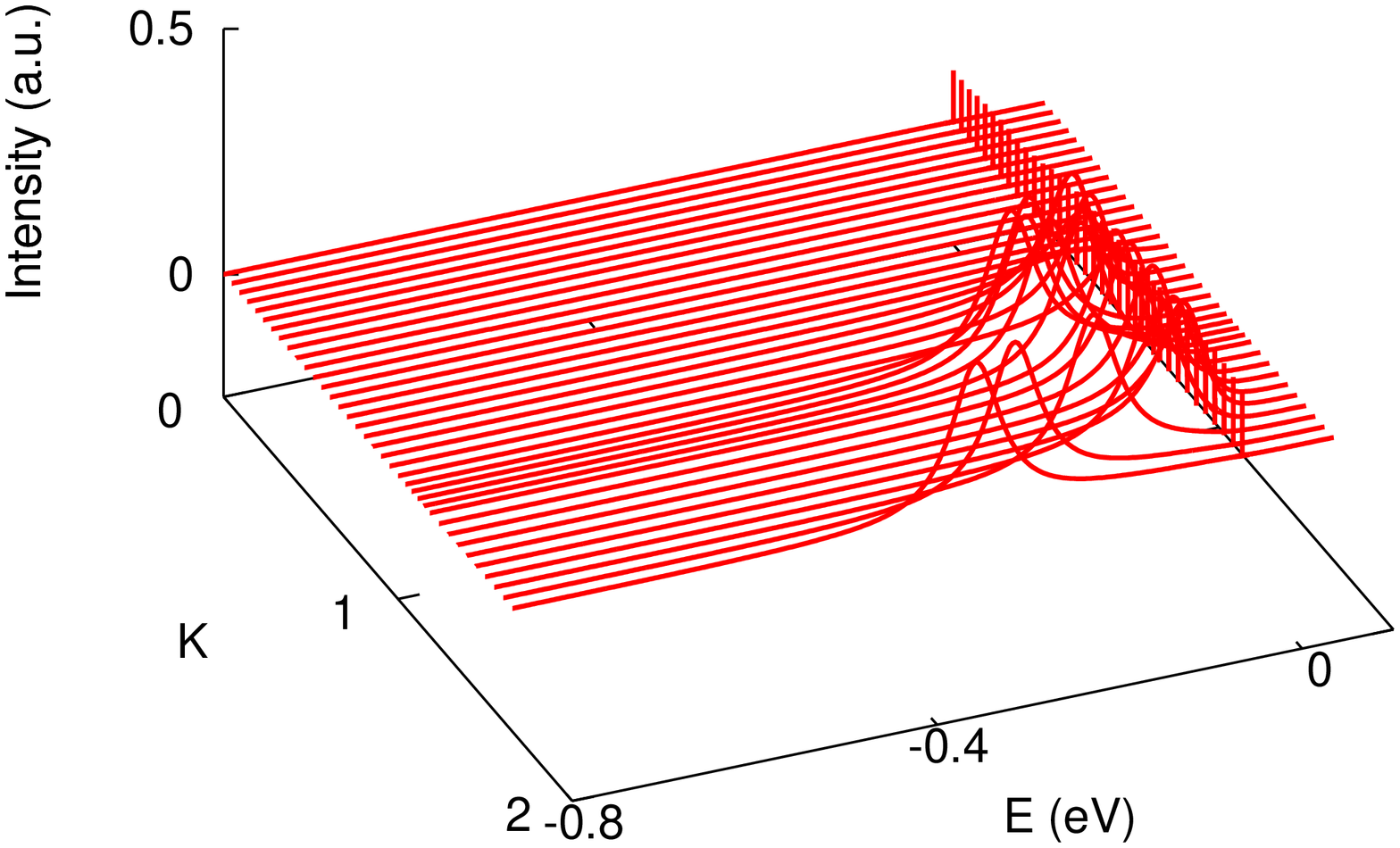}

(g)
\includegraphics[angle=270,width=0.8\columnwidth]{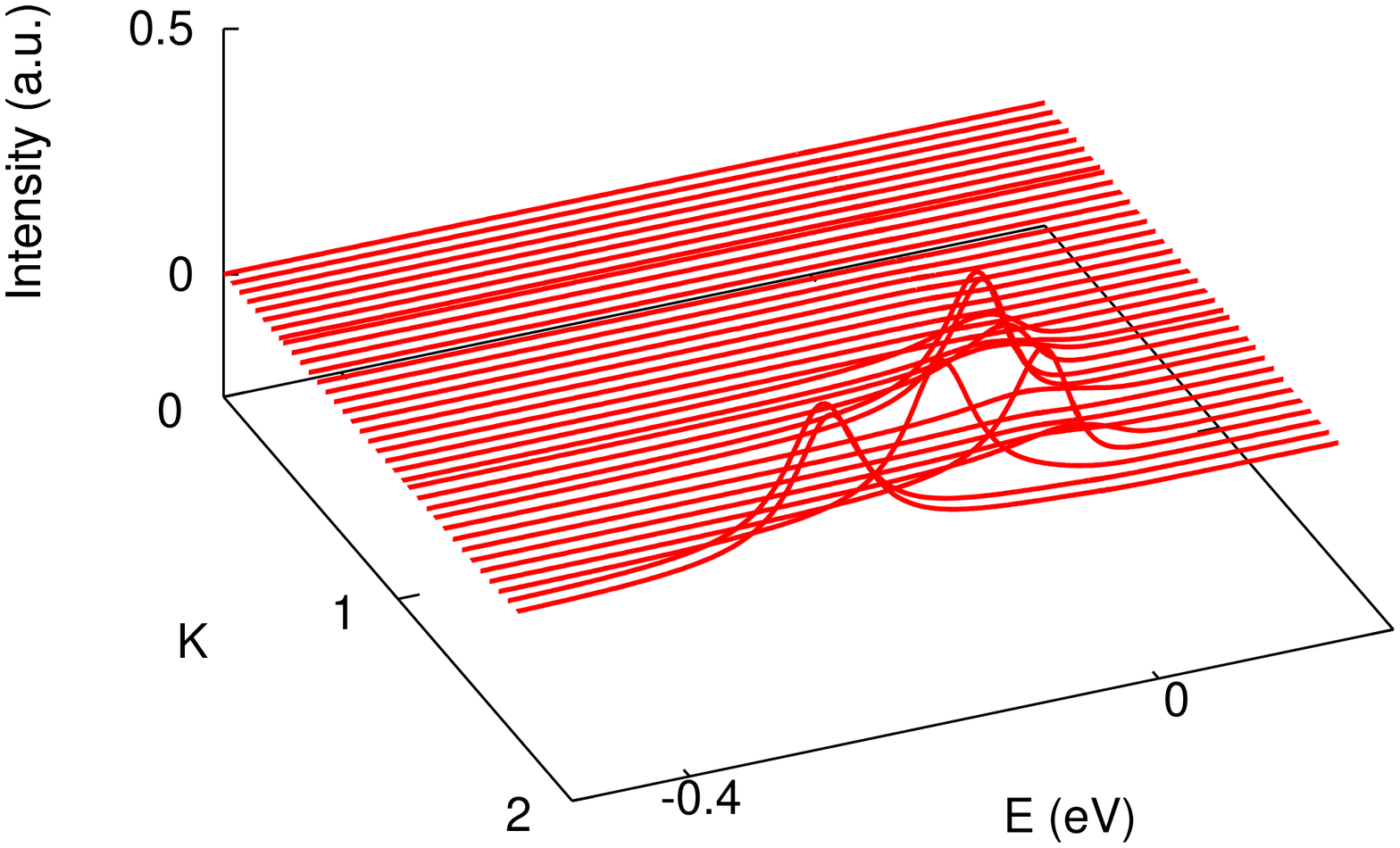}
(h)
\includegraphics[angle=270,width=0.8\columnwidth]{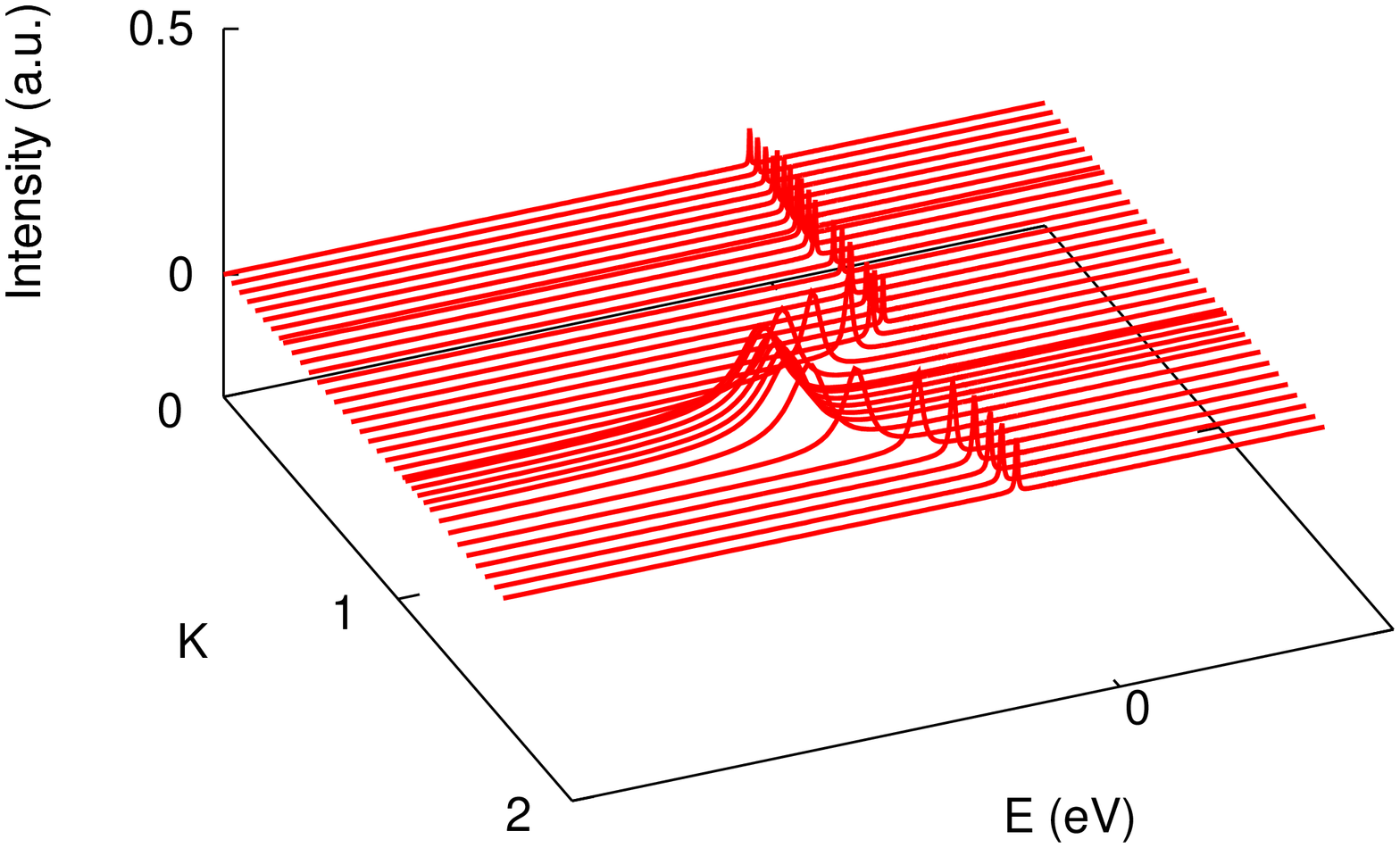}
\caption{(Color Online)DMFT ARPES Intensity of parent $SrFe_2As_2$ at
(a)-(d) 5 K and (e)-(h) 200 K for $xy, x^2-y^2, z^2$ and $yz$ bands along 
$\Gamma-M$ direction.}
\label{figs1}
\end{figure*}

\begin{figure*}
(a)
\includegraphics[angle=270,width=0.8\columnwidth]{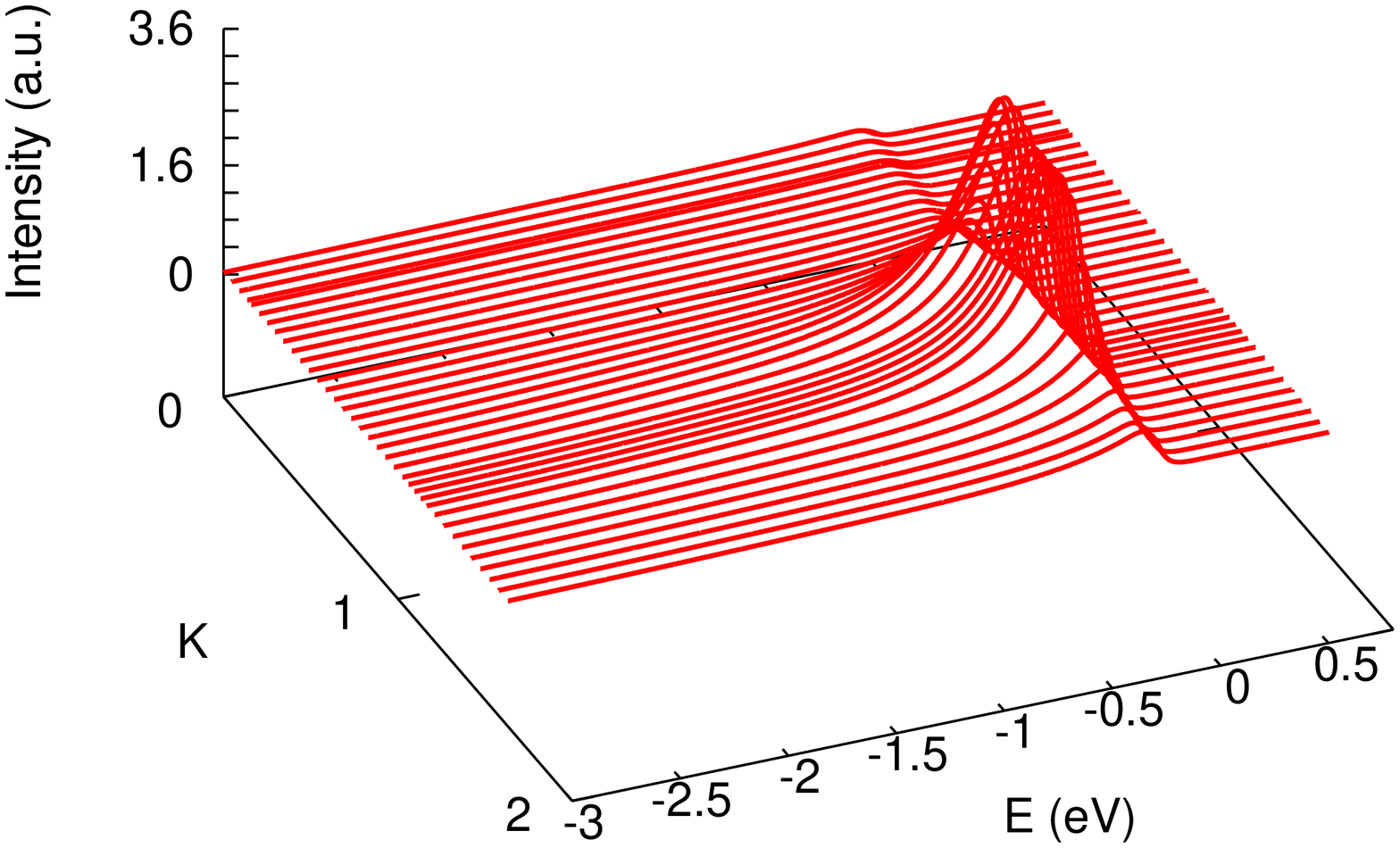}
(b)
\includegraphics[angle=270,width=0.8\columnwidth]{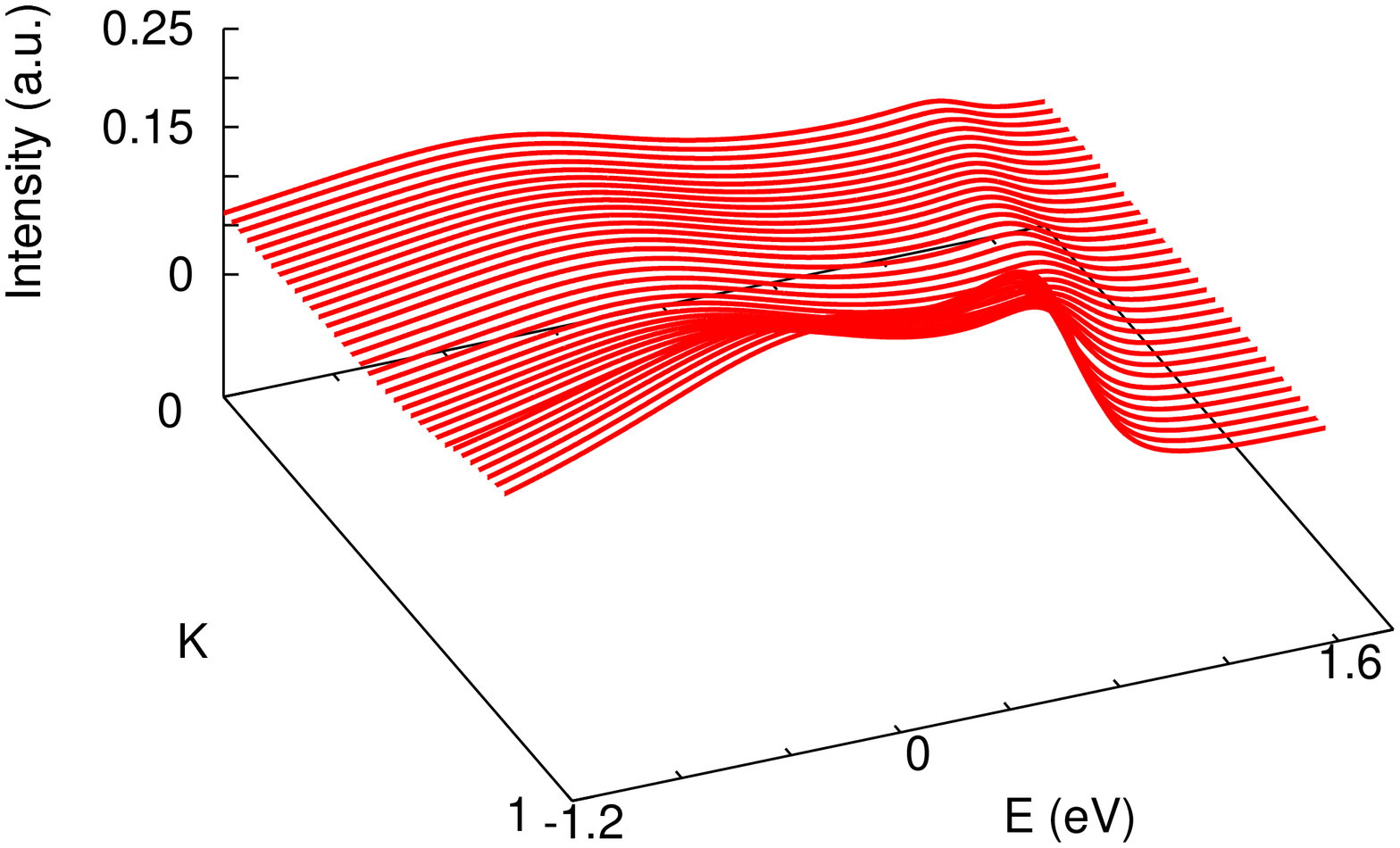}

(c)
\includegraphics[angle=270,width=0.8\columnwidth]{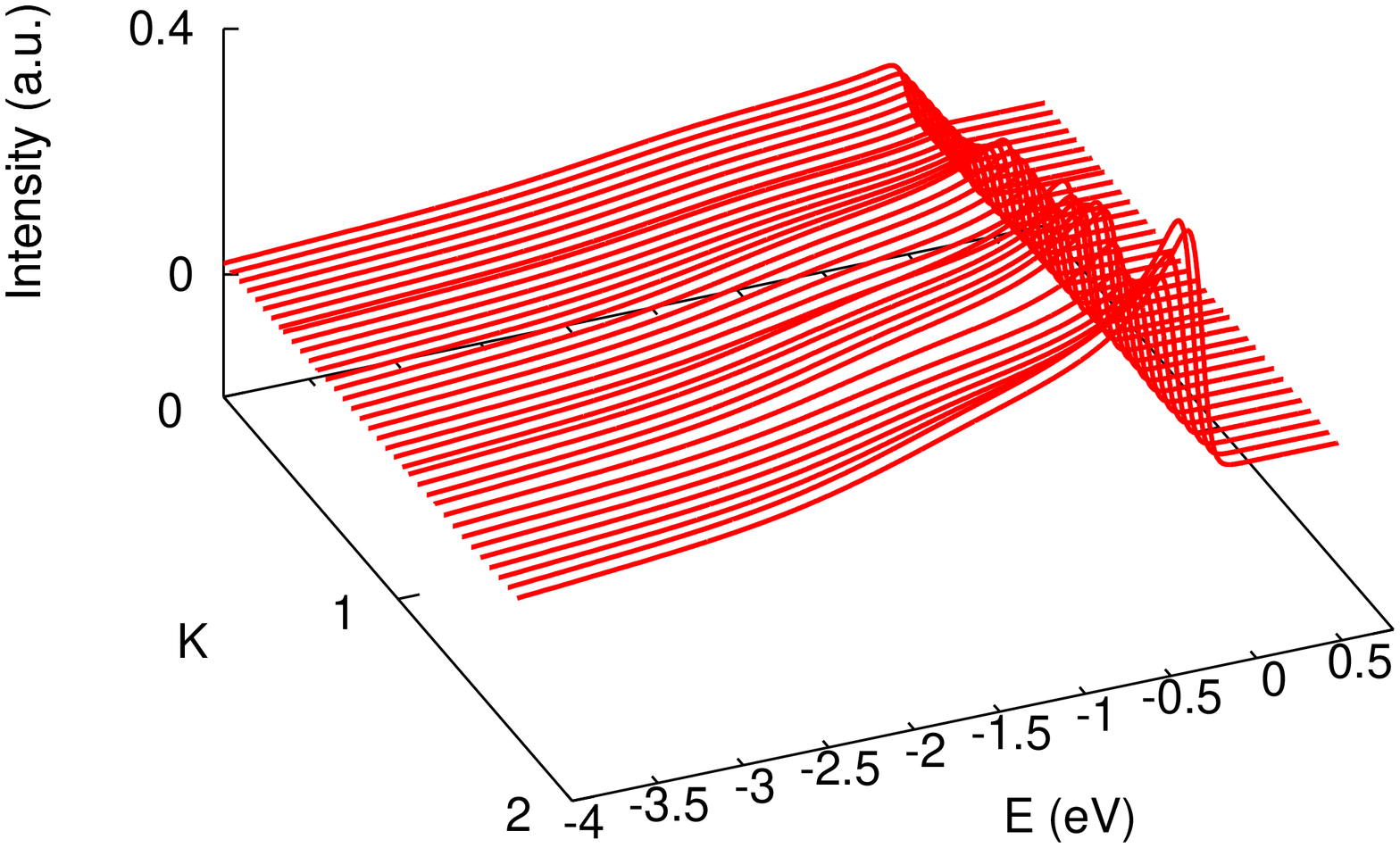}
(d)
\includegraphics[angle=270,width=0.8\columnwidth]{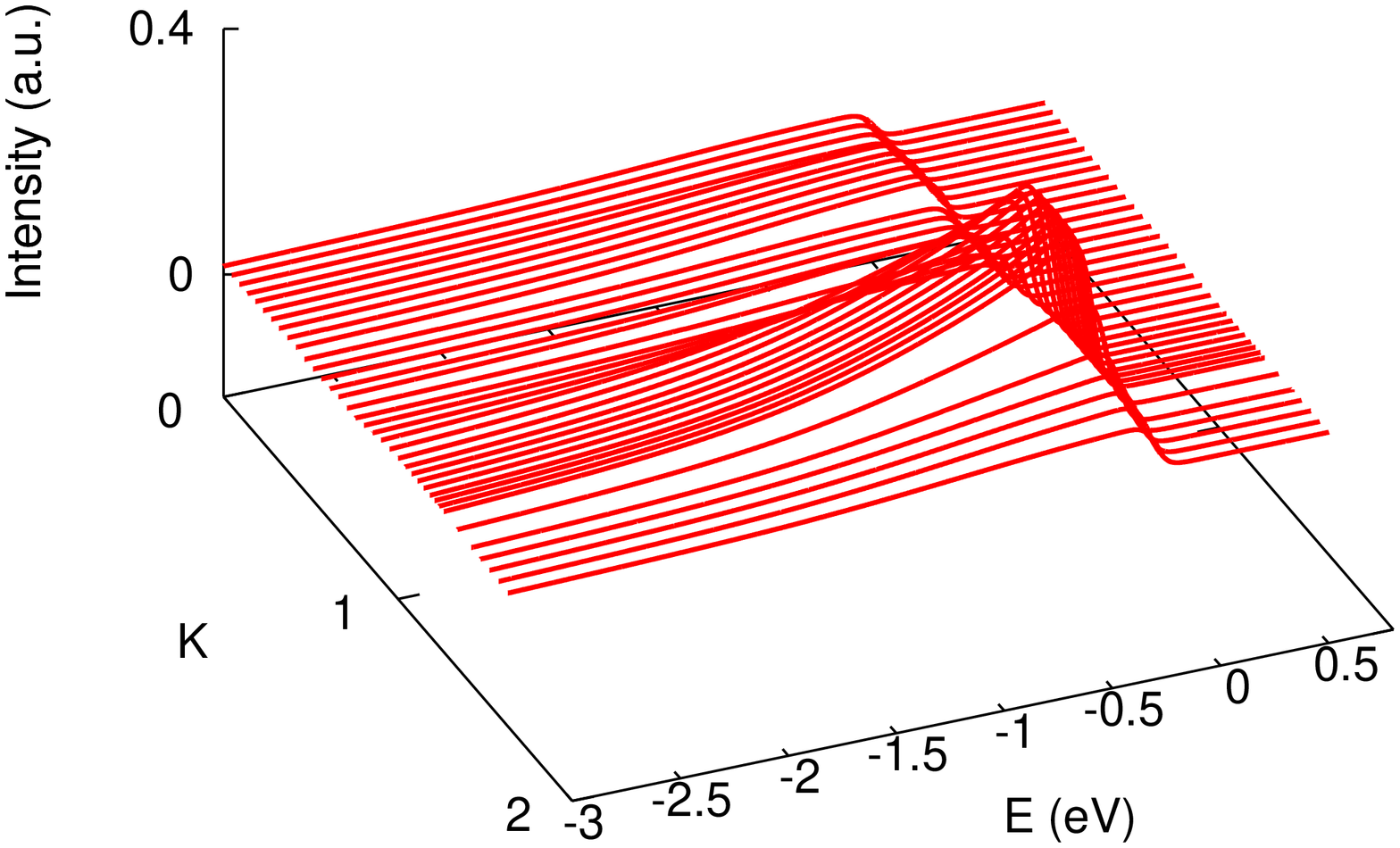}

\caption{(Color Online)(a)-(d) DMFT ARPES Intensity of parent $SrFe_2As_2$ at
300 K for $xy, x^2-y^2, z^2$ and $yz$ bands along $\Gamma-M$ direction.}
\label{figs2}
\end{figure*}

\begin{figure*}
(a)
\includegraphics[angle=270,width=0.8\columnwidth]{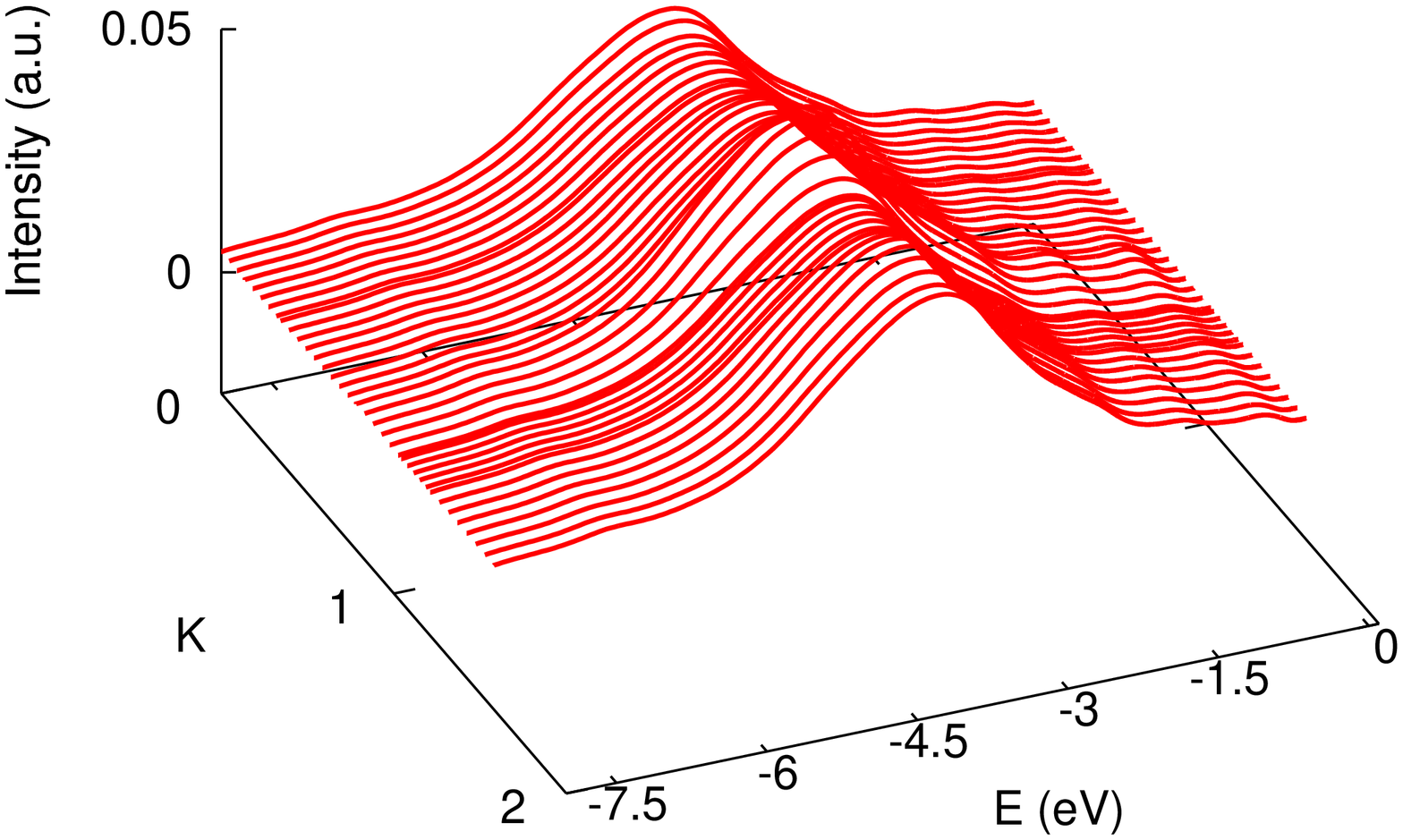}
(b)
\includegraphics[angle=270,width=0.8\columnwidth]{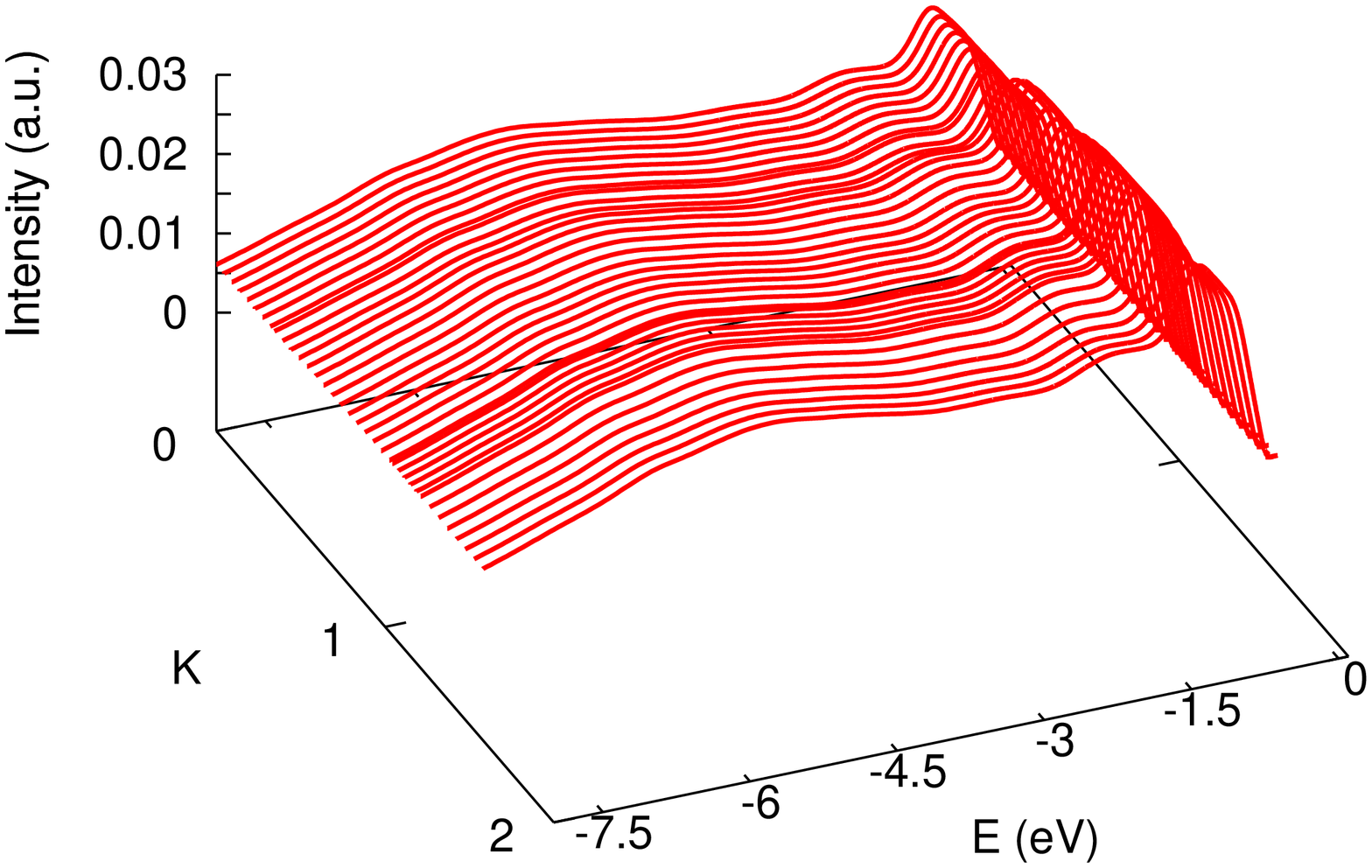}

(c)
\includegraphics[angle=270,width=0.8\columnwidth]{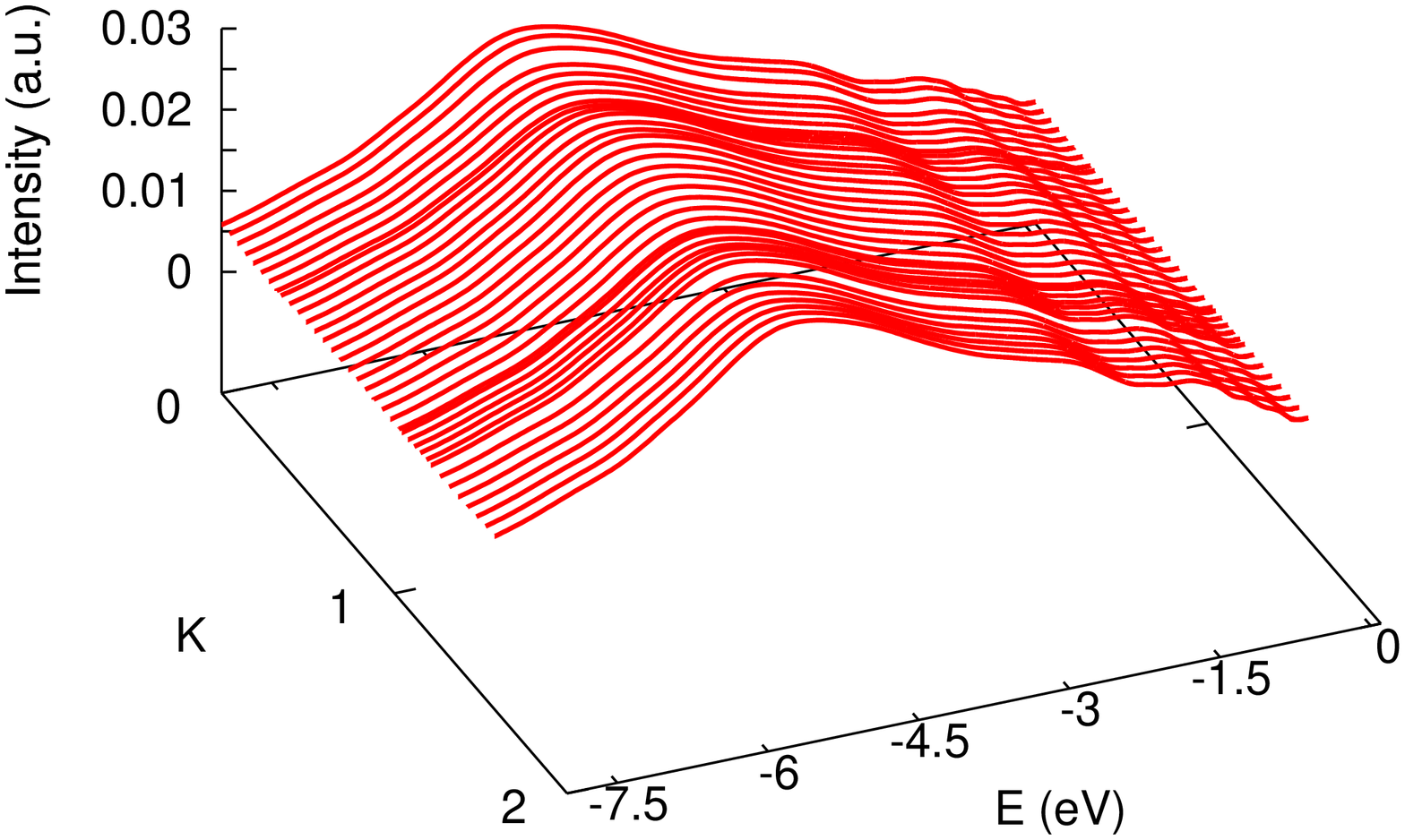}
(d)
\includegraphics[angle=270,width=0.8\columnwidth]{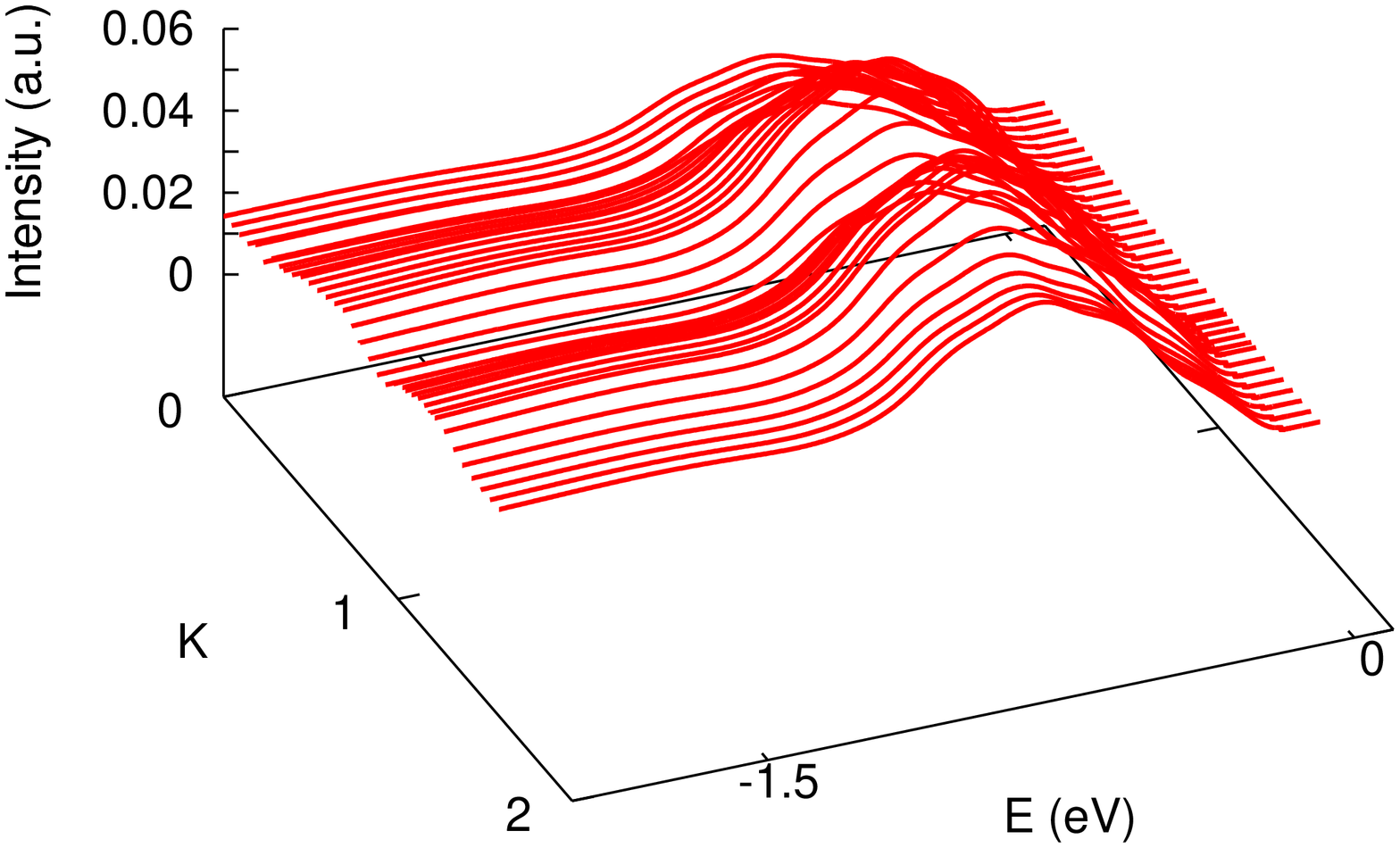}

(e)
\includegraphics[angle=270,width=0.8\columnwidth]{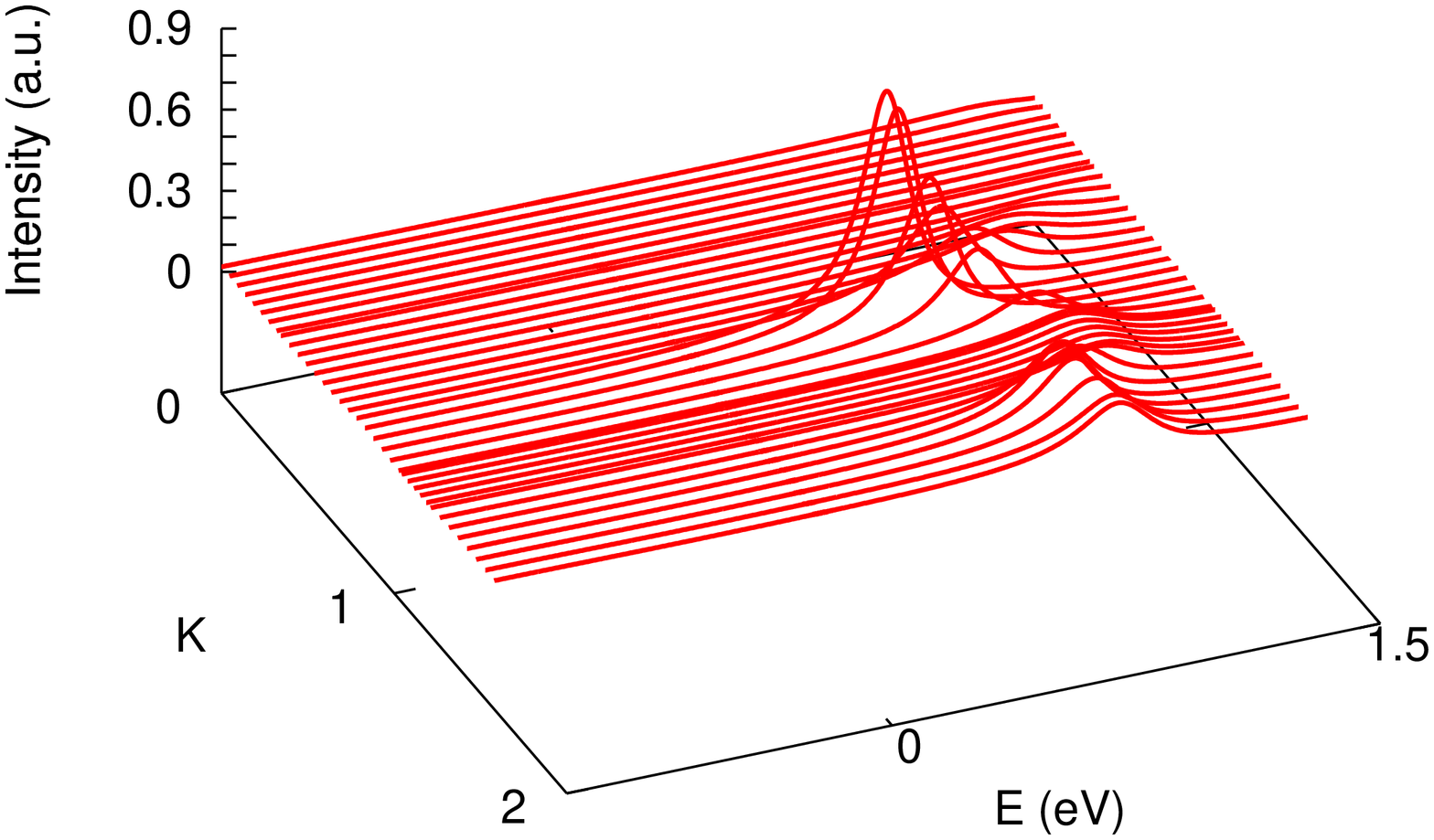}
(f)
\includegraphics[angle=270,width=0.8\columnwidth]{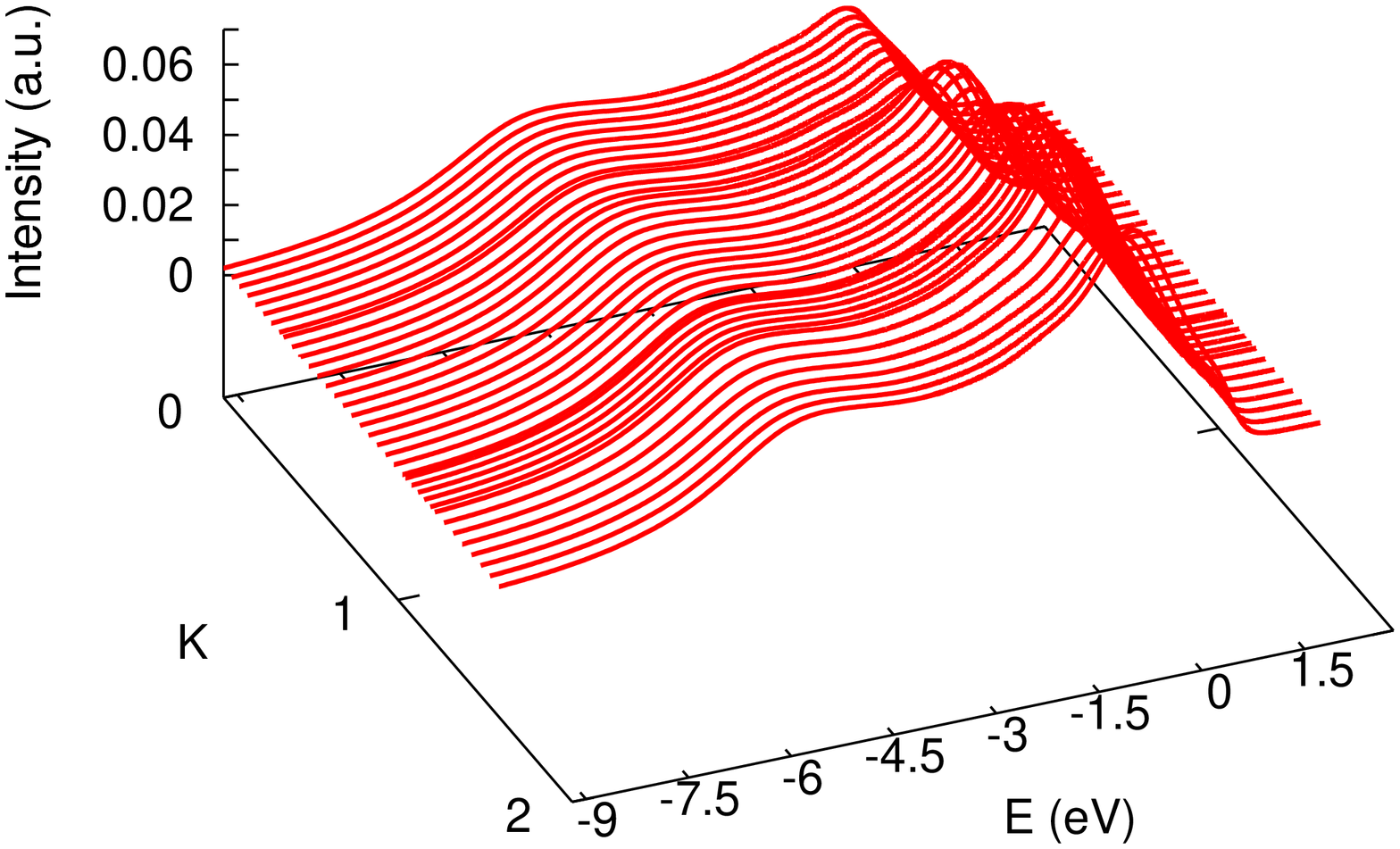}

(g)
\includegraphics[angle=270,width=0.8\columnwidth]{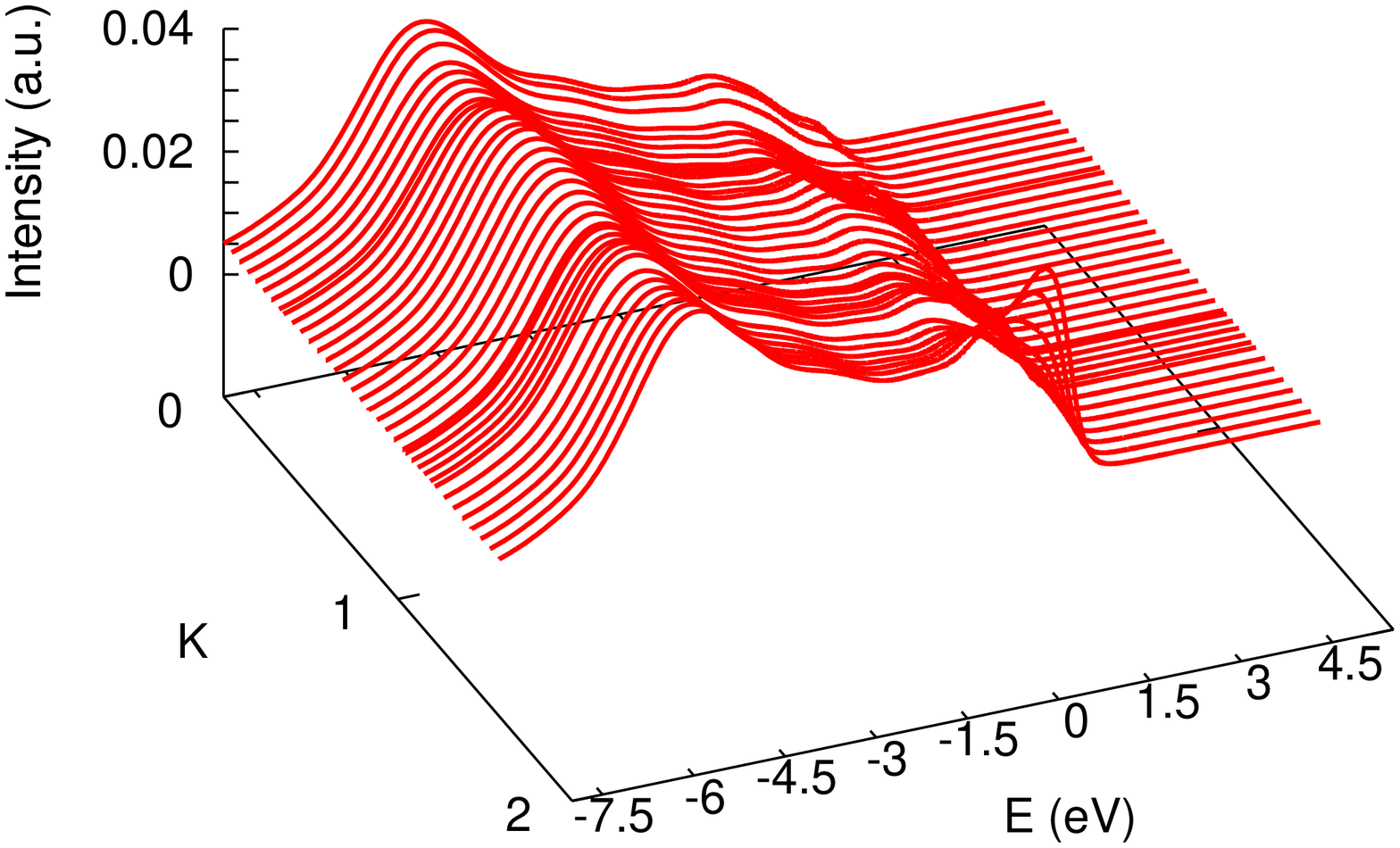}
(h)
\includegraphics[angle=270,width=0.8\columnwidth]{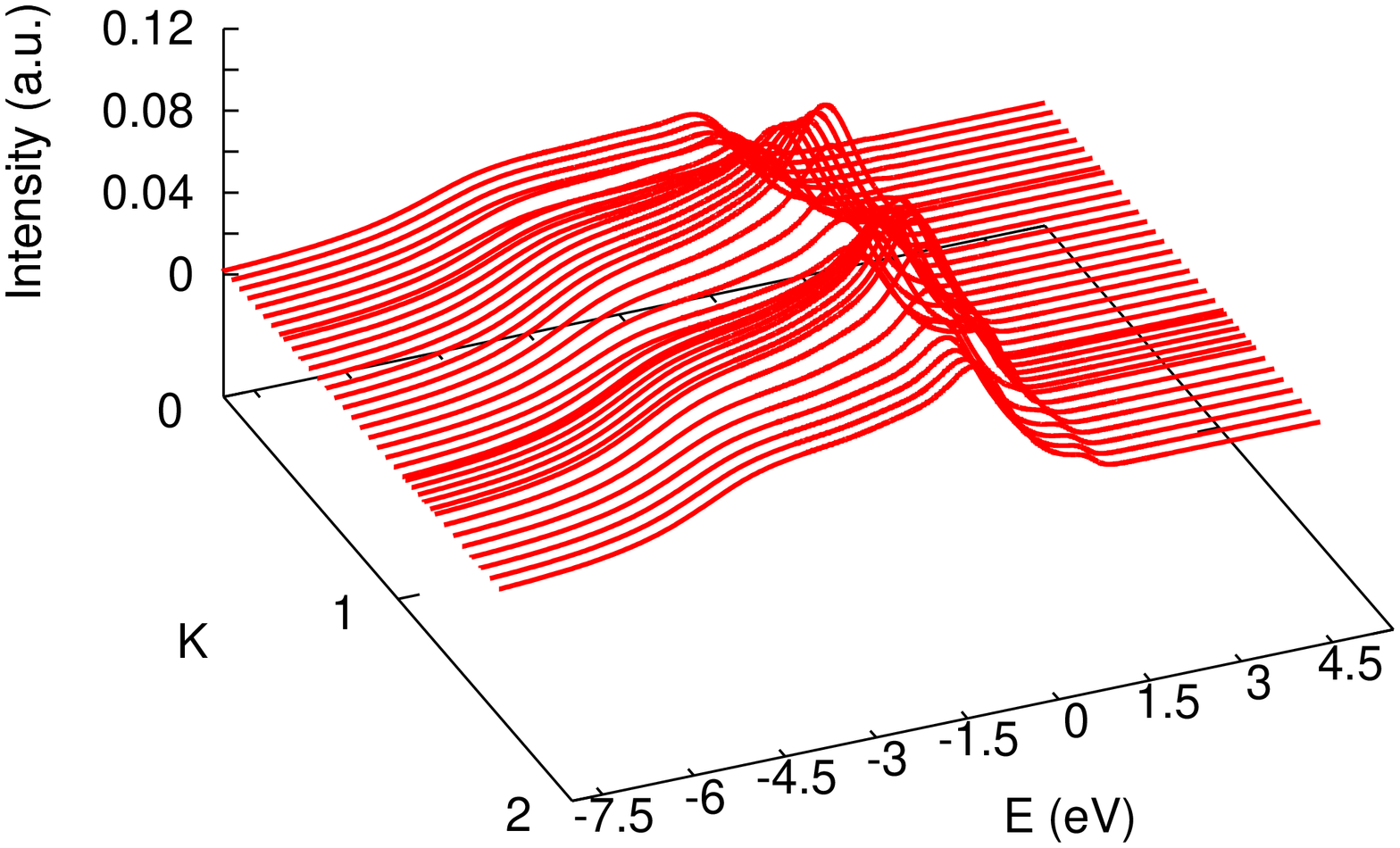}
\caption{(Color Online)DMFT ARPES Intensity of $SrFe_{1.86}Ni_{0.14}As_2$ at
(a)-(d) 150 K and (e)-(h) 300 K for $xy, x^2-y^2, z^2$ and $yz$ bands along $\Gamma-M$ direction.}
\label{figs3}
\end{figure*}

\begin{figure*}
(a)
\includegraphics[angle=270,width=0.8\columnwidth]{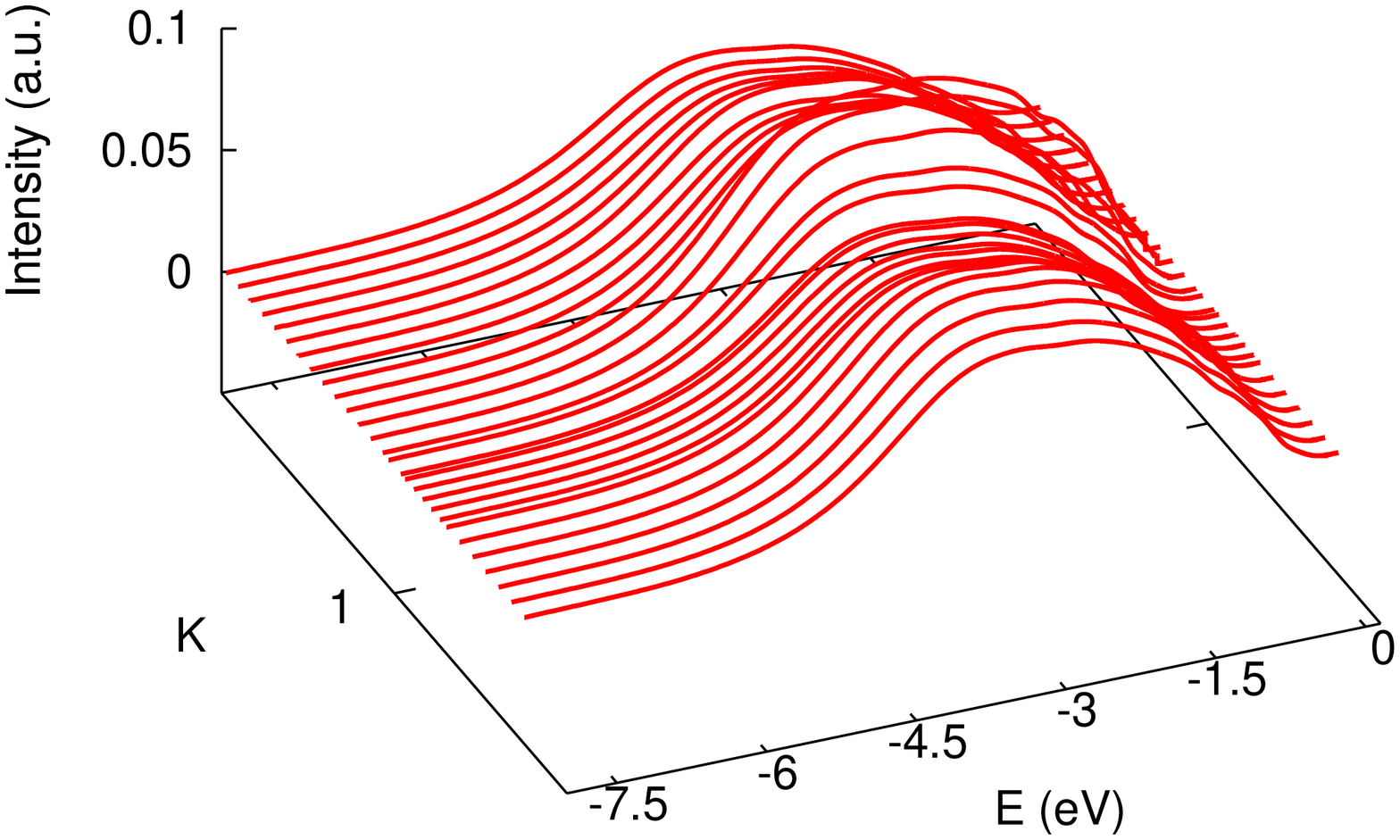}
(b)
\includegraphics[angle=270,width=0.8\columnwidth]{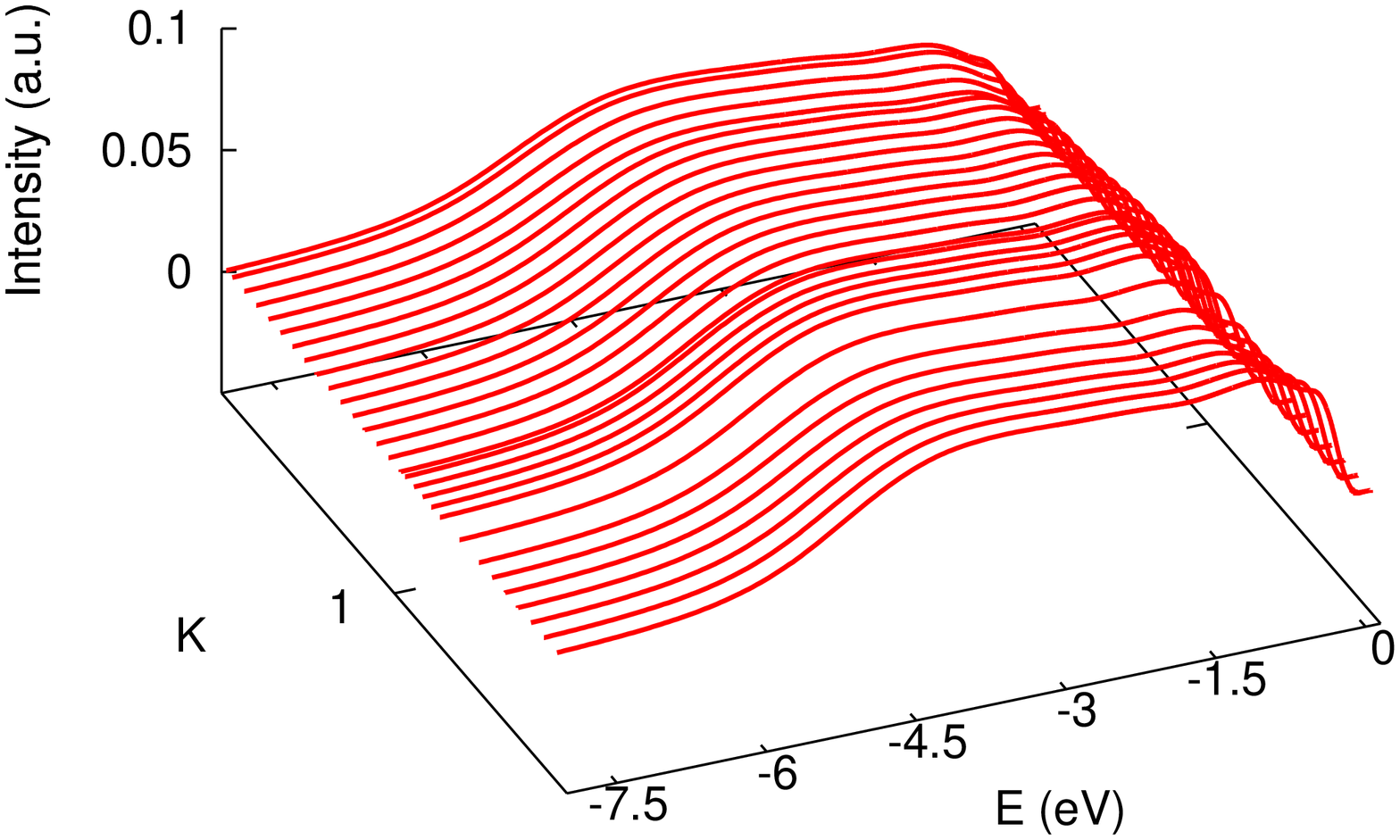}

(c)
\includegraphics[angle=270,width=0.8\columnwidth]{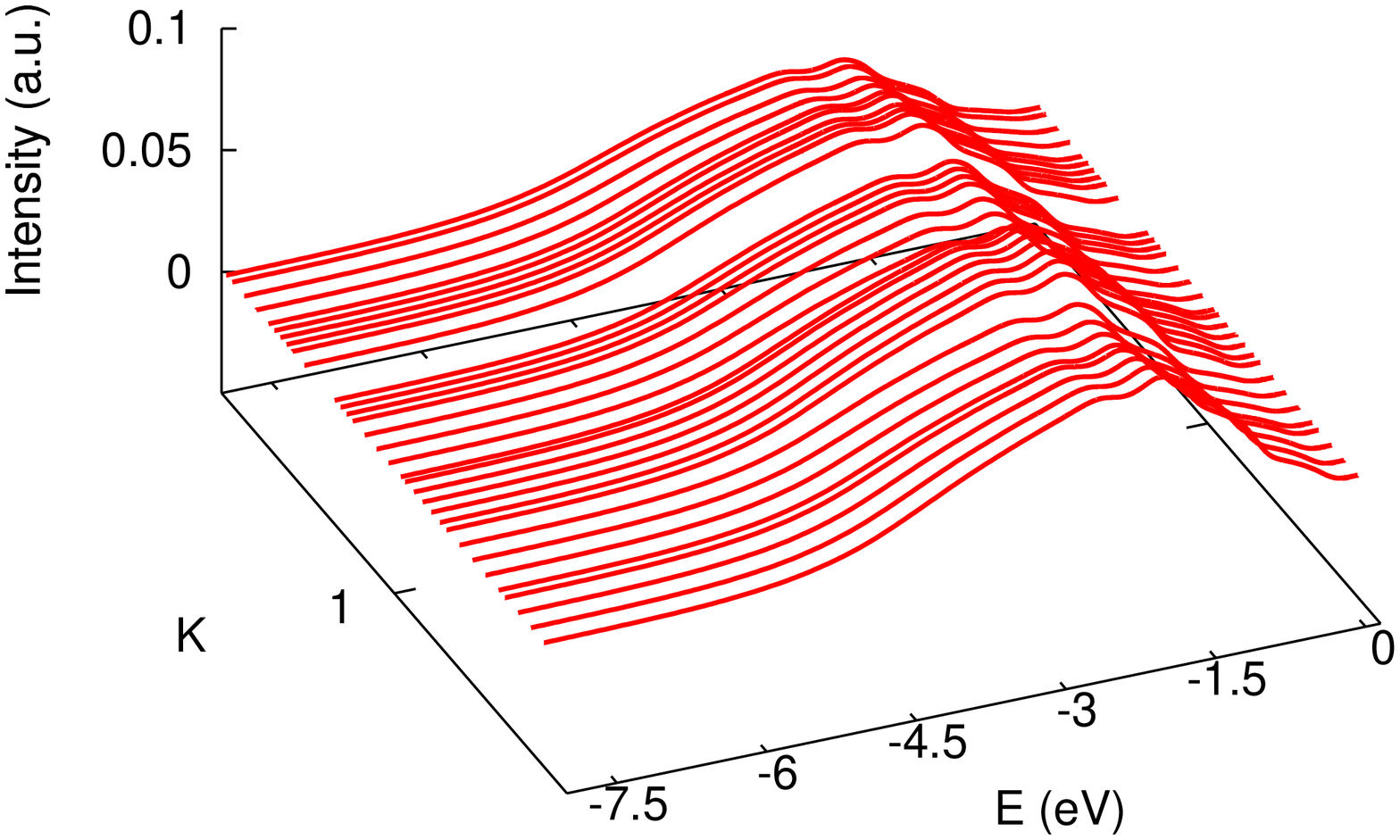}
(d)
\includegraphics[angle=270,width=0.8\columnwidth]{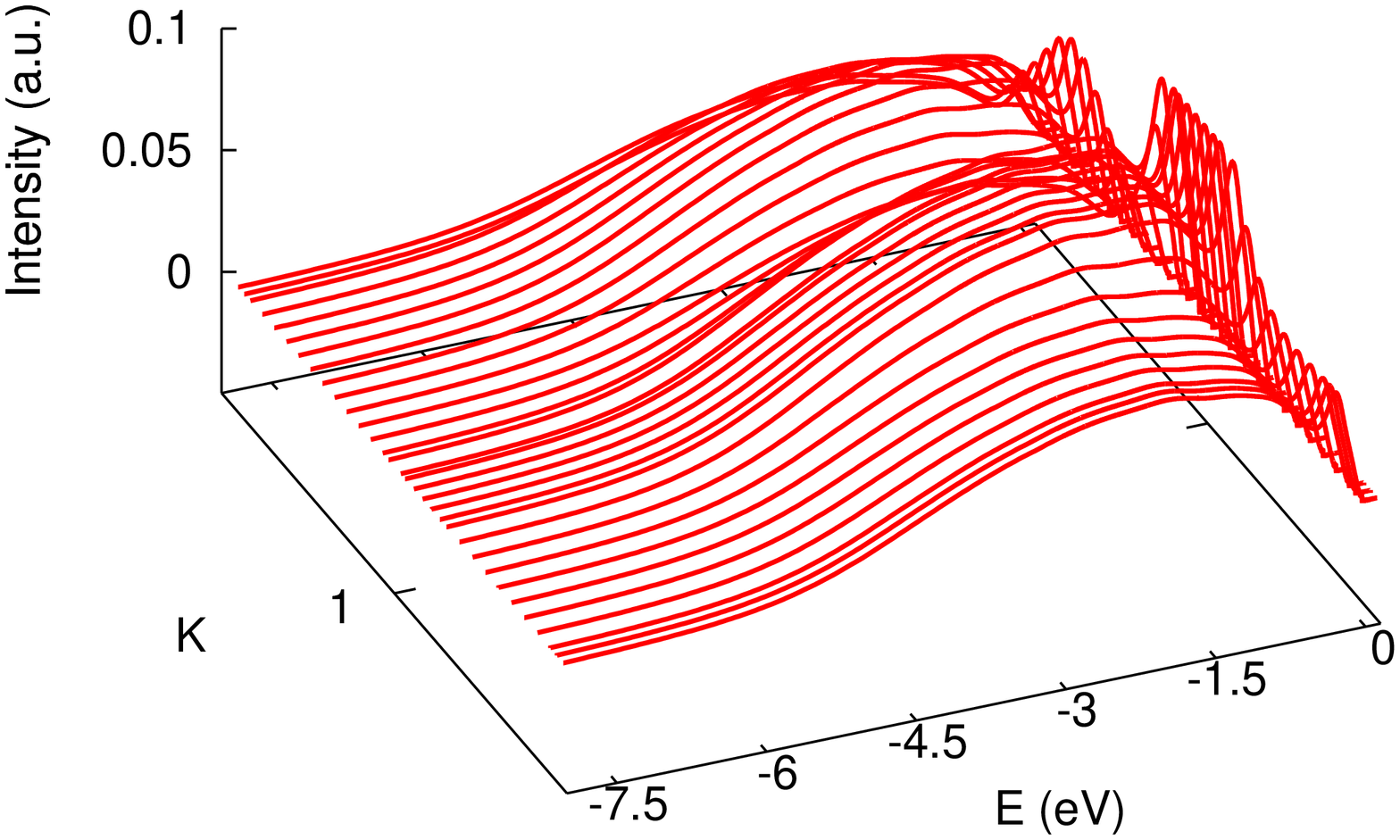}

(e)
\includegraphics[angle=270,width=0.8\columnwidth]{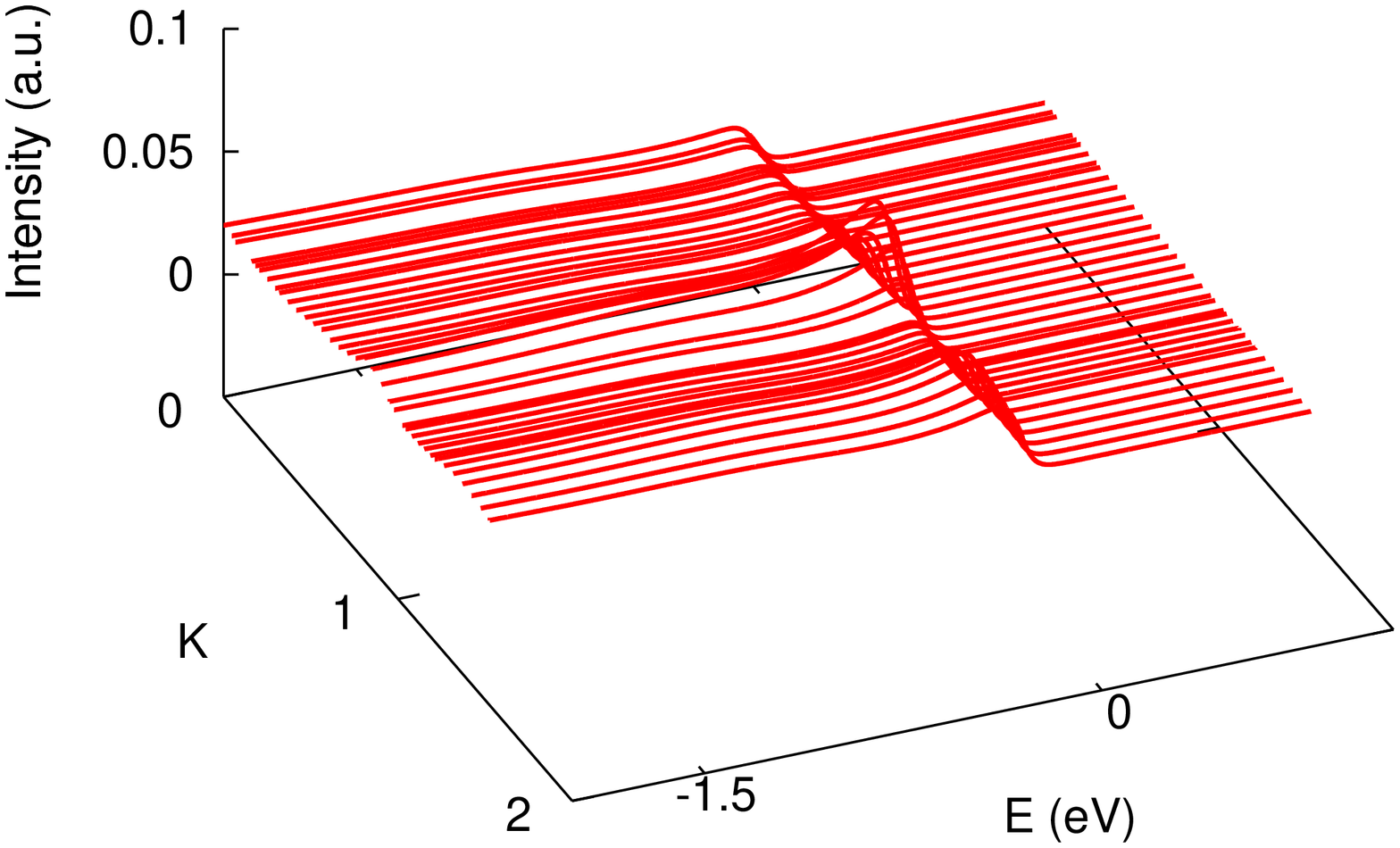}
(f)
\includegraphics[angle=270,width=0.8\columnwidth]{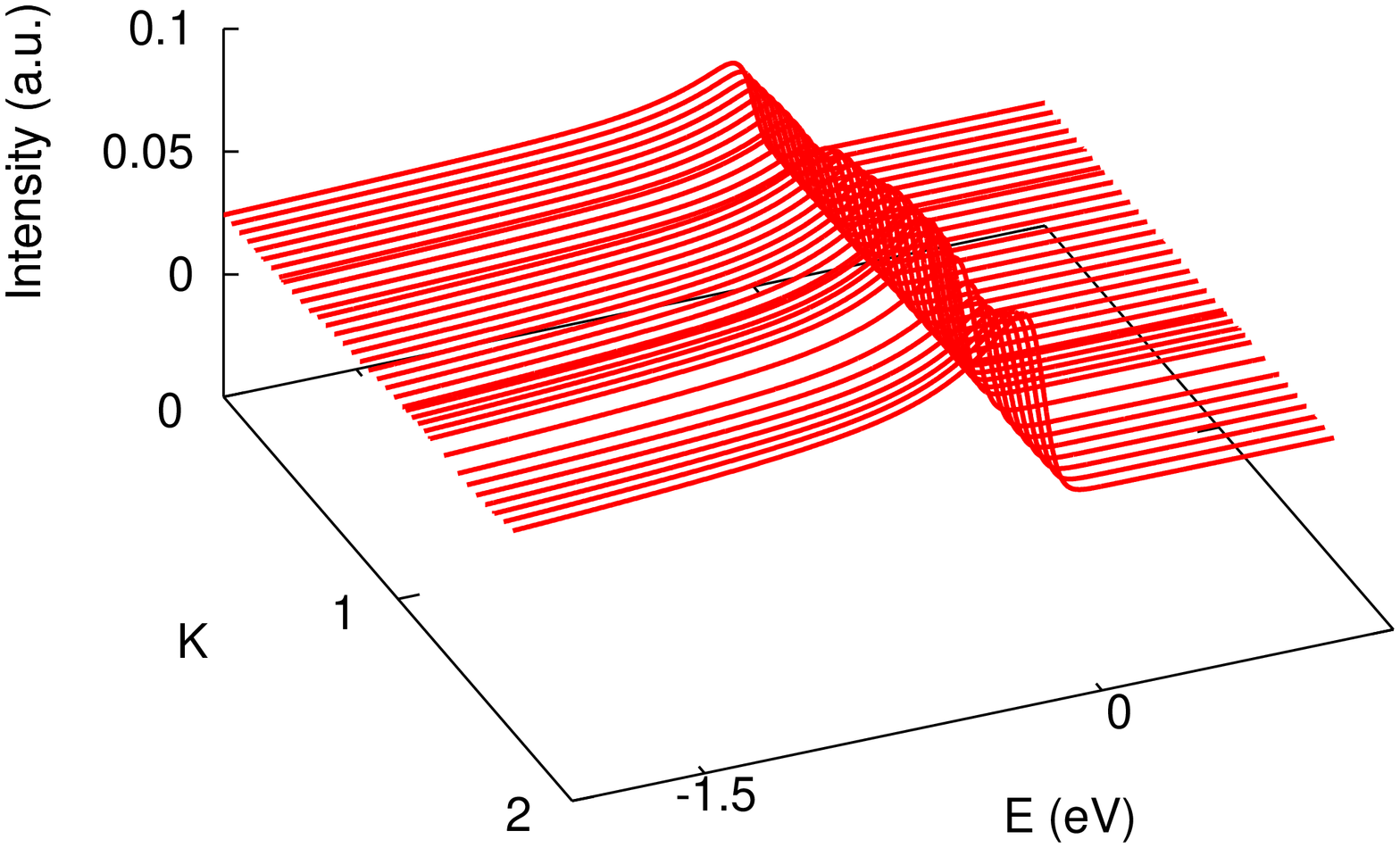}

(g)
\includegraphics[angle=270,width=0.8\columnwidth]{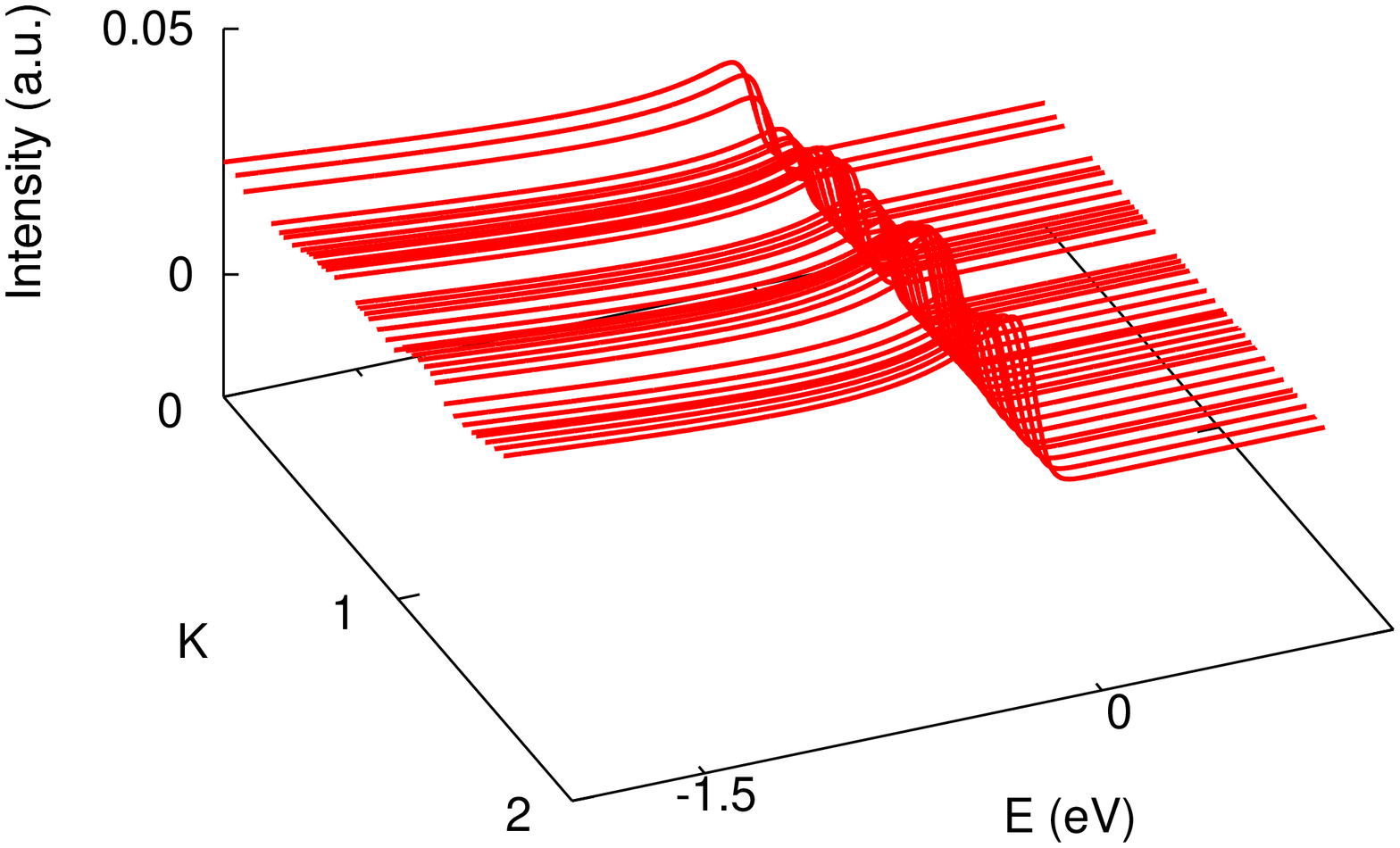}
(h)
\includegraphics[angle=270,width=0.8\columnwidth]{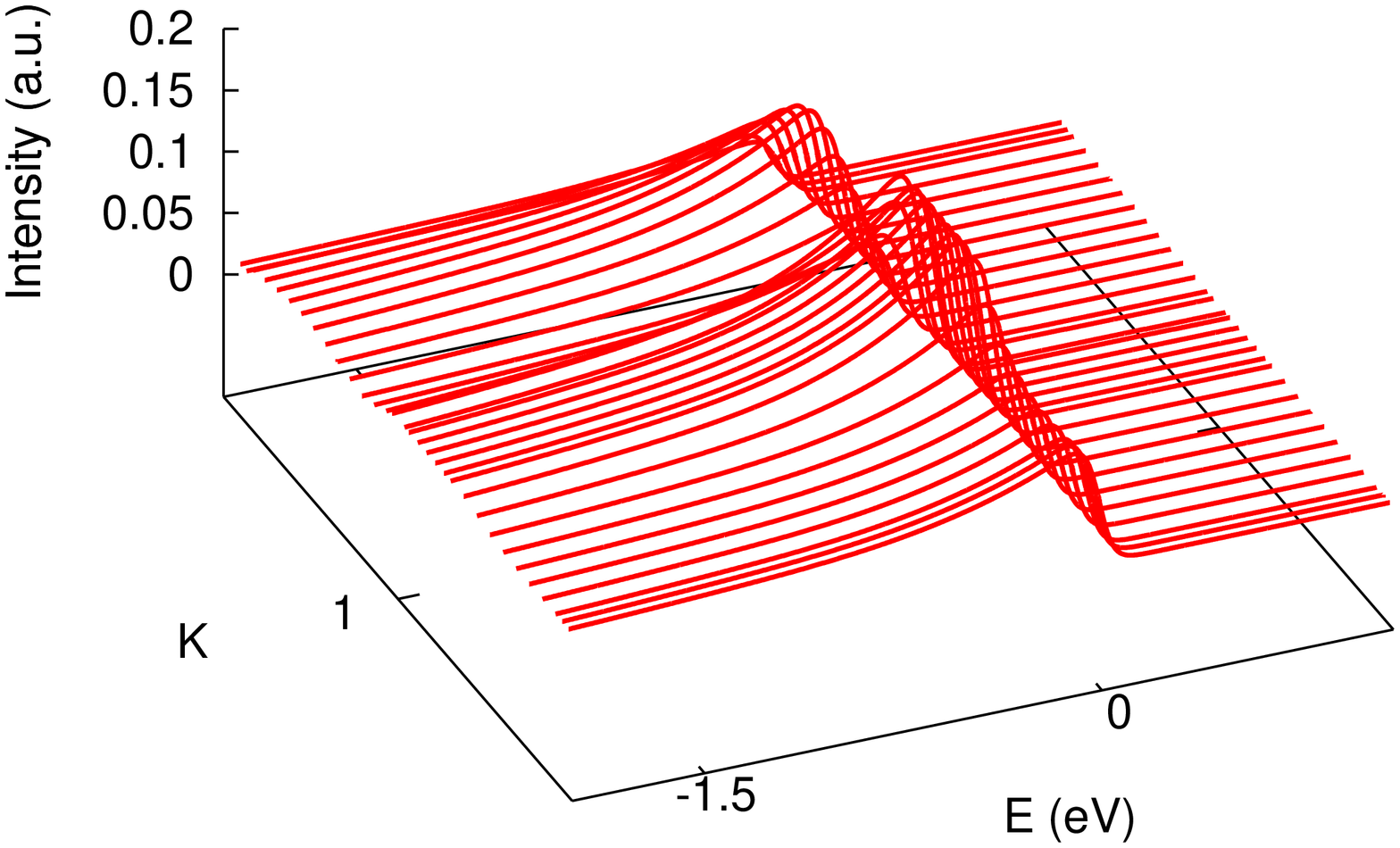}
\caption{(Color Online)DMFT ARPES Intensity of $SrFe_{1.8}Ni_{0.2}As_2$ at
(a)-(d) 150 K and (e)-(h) 300 K for $xy, x^2-y^2, z^2$ and $yz$ bands along $\Gamma-M$ direction.}
\label{figs4}
\end{figure*}
\begin{figure}
\centering
\includegraphics[angle=270,width=0.8\columnwidth]{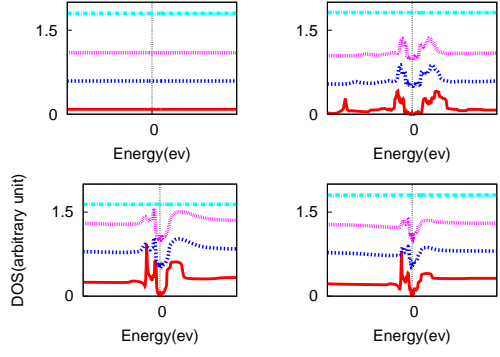}
\caption{(Color Online)DMFT density of states of parent $SrFe_2As_2$ at 
different temperature. In upper pannel $yz$, $x^2-y^2$ and 
in lower panel $xy$ and $z^2$ DOS is shown. Clear gap at $E_F$ manifests 
orbital selective Mott localization.}
\label{figs5}
\end{figure}
\end{document}